\documentclass[11pt]{report}
\usepackage{geometry, graphicx, kotex, imakeidx, titlesec, array} 
\usepackage{mathptmx}   
\usepackage{amsmath, amsthm, amssymb, mathrsfs, multirow, verbatim} 
\usepackage[nobiblatex]{xurl} 
\usepackage{indentfirst} 
\usepackage{booktabs} 
\usepackage{ragged2e}
\usepackage{setspace}
\usepackage{cite}
\usepackage{amsmath,amssymb,amsfonts}
\usepackage{algorithmic}
\usepackage{graphicx}
\usepackage{textcomp}
\usepackage{multirow}
\usepackage{array}
\usepackage{amsmath,amssymb,amsfonts}
\usepackage{graphicx}
\usepackage{textcomp}
\usepackage{multirow}
\usepackage{array}
\usepackage{tikz}
\usepackage{footnote}
\usepackage[ruled,vlined, linesnumbered]{algorithm2e}
\usepackage{rotating}
\usepackage{booktabs}
\usepackage{pdflscape}
\usepackage{longtable}
\usepackage{caption}
\usepackage{float}
\usepackage{rotfloat}

\titleformat{\chapter}{\normalfont\huge\bfseries}{\chaptertitlename\ \thechapter.}{20pt}{\huge} 
\makeindex 

\renewcommand\arraystretch{1.3} 

\begin{document}

\newpage
\thispagestyle{empty}
\begin{center}
\noindent
{\fontsize{16pt}{16pt}\selectfont Doctoral Dissertation \par }  
\vspace{3cm} 
\LARGE A Quantitative Framework for Assessing Sleep Quality from EEG Time Series in Complex Dynamic Systems 
\par\vspace{4cm} 
{\fontsize{16pt}{16pt}\selectfont Gi-Hwan Shin \par}
\vspace{0.5cm}
{\fontsize{16pt}{16pt}\selectfont Department of Brain and Cognitive Engineering \par} 
\vspace{1.5cm}
{\fontsize{18pt}{18pt}\selectfont Graduate School \par} 
\vspace{0.5cm}
{\fontsize{18pt}{18pt}\selectfont Korea University \par}
\vspace{1cm}
{\fontsize{14pt}{14pt}\selectfont February 2026} 
\end{center}


\newpage
\thispagestyle{empty}
\begin{center}
\LARGE A Quantitative Framework for Assessing Sleep Quality from EEG Time Series in Complex Dynamic Systems 
\par\vspace{1.5cm} 
{\fontsize{16pt}{16pt}\selectfont by \par Gi-Hwan Shin \par} 
\vspace{0.5cm}
\rule{.6\textwidth}{0.4pt} 
\par\vspace{0.2cm}
{\fontsize{16pt}{18pt}\selectfont under the supervision of \par Professor Seong-Whan Lee \par 
\vspace{0.7cm}
A dissertation submitted in partial fulfillment of \par
the requirements for the degree of \par
Doctor of Philosophy \par } 
\vspace{10pt}
{\fontsize{16pt}{16pt}\selectfont Department of Brain and Cognitive Engineering \par }  
\vspace{1.5cm}
{\fontsize{18pt}{18pt}\selectfont Graduate School \par\vspace{0.2cm}
 Korea University \par} 
\vspace{1cm}
{\fontsize{14pt}{14pt}\selectfont February 2026} 
\end{center}


\newpage
\thispagestyle{empty}
\begin{center}
\vspace{1cm}
{\fontsize{16pt}{18pt}\selectfont
The dissertation of Gi-Hwan Shin has been approved by the dissertation committee in partial fulfillment 
of the requirements for the degree of \par
Doctor of Philosophy \par}
\vspace{1cm}
{\fontsize{14pt}{14pt}\selectfont  February 2026 \par}  
\vspace{2cm}
\rule{.6\textwidth}{0.4pt}\par 
{\fontsize{16pt}{16pt}\selectfont Committee Chair: Seong-Whan Lee \par}
\vspace{1cm}
\rule{.6\textwidth}{0.4pt}\par 
{\fontsize{16pt}{16pt}\selectfont Committee Member: Wonzoo Chung \par}
\vspace{1cm}
\rule{.6\textwidth}{0.4pt}\par 
{\fontsize{16pt}{16pt}\selectfont Committee Member: Tae-Eui Kam \par}
\vspace{1cm}
\rule{.6\textwidth}{0.4pt}\par 
{\fontsize{16pt}{16pt}\selectfont Committee Member: Jichai Jeong \par}
\vspace{1cm}
\rule{.6\textwidth}{0.4pt}\par 
{\fontsize{16pt}{16pt}\selectfont Committee Member: Sanghoon Sull \par}
\vspace{1cm}
\end{center}

\newpage
\pagenumbering{roman} 
\newgeometry{left=20mm, right=20mm, top=30mm, bottom=30mm} 
\begin{center}
\LARGE A Quantitative Framework for Assessing Sleep Quality from EEG Time Series in Complex Dynamic Systems 
\par\vspace{20pt}

\normalsize \doublespacing
by Gi-Hwan Shin\par 
Department of Brain and Cognitive Engineering\par
under the supervision of Professor Seong-Whan Lee 
\par\vspace{20pt}
\large \textbf{Abstract}
\end{center}

\normalsize
\justifying 
\doublespacing

Modern lifestyles contribute to insufficient sleep, impairing cognitive function and weakening the immune system. Sleep quality (SQ) is vital for physiological and mental health, making its understanding and accurate assessment critical. However, its multifaceted nature, shaped by neurological and environmental factors, makes precise quantification challenging. Here, we address this challenge by utilizing electroencephalography (EEG) for phase-amplitude coupling (PAC) analysis to elucidate the neurological basis of SQ, examining both states of sleep and wakefulness, including resting state (RS) and working memory. Our results revealed distinct patterns in beta power and delta connectivity in sleep and RS, together with the reaction time of working memory. A notable finding was the pronounced delta-beta PAC, a feature markedly stronger in individuals with good SQ. We further observed that SQ was positively correlated with increased delta-beta PAC. Leveraging these insights, we applied machine learning models to classify SQ at an individual level, demonstrating that the delta-beta PAC outperformed other EEG characteristics. These findings establish delta-beta PAC as a robust electrophysiological marker to quantify SQ and elucidate its neurological determinants. 

\par\vspace{20pt}


\newpage 
\begin{center}
\LARGE 복잡한 동적 시스템에서 뇌파 시계열로부터 수면의 질 평가를 위한 정량적 프레임워크 
\par\vspace{10pt}
\normalsize 신 기 환\par 
뇌 공 학 과\par 
지 도 교 수:  이 성 환
\par\vspace{10pt}
\large \textbf{국문 초록}
\end{center}

\normalsize 

현대인의 생활 방식은 불충분한 수면을 초래하며, 이는 인지 기능을 저하시키고 면역 체계를 약화시킵니다. 수면의 질은 생리적 및 정신적 건강에 필수적이므로, 이를 이해하고 정확하게 평가하는 것은 매우 중요합니다. 그러나 신경학적 및 환경적 요인에 의해 형성되는 수면의 질의 다면적 특성으로 인해 정확한 정량화가 어렵습니다. 본 연구에서는 뇌파를 이용한 위상-진폭 결합 분석을 통해 수면의 질의 신경학적 기전을 규명하고, 휴지기 상태와 작업 기억을 포함한 수면 및 각성 상태를 모두 조사함으로써 이러한 문제를 해결하고자 합니다. 연구 결과, 작업 기억의 반응 시간과 더불어 수면 및 휴지기 상태에서의 베타 파워와 델타 연결성에서 뚜렷한 패턴이 나타났습니다. 주목할 만한 발견은 두드러진 델타-베타 위상-진폭 결합이었으며, 이 특징은 수면의 질이 좋은 개인에게서 상당히 강하게 나타났습니다. 우리는 나아가 수면의 질이 델타-베타 위상-진폭 결합의 증가와 양의 상관관계가 있음을 관찰했습니다. 이러한 통찰력을 활용하여, 우리는 개인 수준에서 수면의 질을 분류하기 위해 머신러닝 모델을 적용하였으며, 델타-베타 위상-진폭 결합이 다른 뇌파 특성들보다 우수한 성능을 보임을 입증했습니다. 이러한 결과는 델타-베타 위상-진폭 결합이 수면의 질을 정량화하고 그 신경학적 결정 요인을 규명하기 위한 강력한 전기생리학적 지표임을 확립합니다.

\par \vspace{10pt}
\textbf{중심어}: 위상-진폭 결합, 수면의 질, 수면, 각성, 뇌파

\newpage
\chapter*{Preface}
\normalsize

This dissertation is submitted for the degree of Doctor of Philosophy in the Department of Brain and Cognitive Engineering at Korea University. The research presented herein was conducted under the supervision of Professor Seong-Whan Lee in the Department of Artificial Intelligence at Korea University. The core content of this dissertation is based on the study published in \textit{IEEE Transactions on Neural Systems and Rehabilitation Engineering}, volume 33, 2025. Neither this dissertation, nor any substantially similar work, has been or is being submitted for any other degree, diploma, or academic qualification at any other university.

\newpage
\chapter*{Acknowledgment}

This work was partly supported by Institute of Information \& Communications Technology Planning \& Evaluation (IITP) grant funded by the Korea government (MSIT) (No. RS-2019-II190079, Artificial Intelligence Graduate School Program (Korea University)) and the National Research Foundation of Korea (NRF) grant funded by the MSIT (No. 2022-2-00975, MetaSkin: Developing Next-generation Neurohaptic Interface Technology that enables Communication and Control in Metaverse by Skin Touch).

\renewcommand*\contentsname{Contents}
\tableofcontents



\listoffigures


\listoftables




\chapter{Introduction}
\pagenumbering{arabic} 
Sleep quality (SQ) plays a critical role in maintaining cognitive function, emotional well-being, and physical health \cite{pilcher1997sleep}. Modern lifestyles, characterized by insufficient sleep, often lead to negative consequences such as impaired cognitive performance and a weakened immune system \cite{garbarino2021role}. However, quantifying SQ is complex due to its dependence on a variety of factors, including physiological and environmental elements, as well as the impact of events occurring before, during, and after the sleep period \cite{pallos2007quality}. The subjective assessments, such as the Pittsburgh Sleep Quality Index (PSQI), are commonly used to measure SQ but are often limited in scope and unable to fully capture the physiological aspects of sleep \cite{buysse1989pittsburgh}. This complexity necessitates exploring objective evaluation that provides deeper insights into SQ.

Recent advancements in pattern recognition have demonstrated exceptional precision in extracting meaningful features from complex datasets. These capabilities range from attentional mechanisms for small object detection \cite{min2022attentional} and accurate object contour tracking to robust text extraction \cite{roh2007accurate} and automatic parsing in compressed video \cite{lim2000text, lee2001automatic}. Building on these rigorous quantitative frameworks established in high-dimensional data analysis, this study aims to apply such robust analytical approaches to physiological signals for clearer SQ assessment.

One effective way to objectively assess SQ is through electroencephalography (EEG), a fundamental tool in neurophysiological research that provides insights into the intricate neural patterns of brain activity \cite{lee2023seriessleepnet}. EEG analysis typically focuses on two essential components: amplitude and phase \cite{wang2019consistency}. Amplitude reflects neural synchrony and activity intensity, while phase captures the synchronization across neural networks, highlighting brain region connectivity \cite{varela2001brainweb}. By integrating these components, phase-amplitude coupling (PAC) emerges as a critical approach, revealing interactions between low-frequency phase and high-frequency amplitude that shape neural coordination \cite{shi2018cross}.

For accurate PAC analysis, it is essential to select amplitude and phase characteristics that reflect brain dynamics \cite{lee2020frontal}. In amplitude characteristics, the spectrogram method is primarily used to understand the variability and intensity of neural activations across frequency bands \cite{lendner2020electrophysiological}. For phase characteristics, the weighted phase lag index (wPLI) is instrumental in evaluating synchronizing activities across different brain regions \cite{bakhshayesh2019detecting}. Integrating these detailed analyses of amplitude and phase characteristics into PAC allows for a more precise understanding of the interactions between brain rhythms \cite{samiee2017time}. Integrating objective and subjective measures bridges the gap between data and experiences, offering a comprehensive framework for understanding SQ \cite{mccarter2022physiological}.

To achieve a deeper understanding of SQ, a comprehensive approach that examines both sleep and wakefulness states is essential \cite{ramlee2017sways}. In the sleep state, characterized by distinct stages such as rapid eye movement (REM) and non-REM (including N1, N2, and N3) \cite{berry2012aasm}, unique neural activities and connectivity patterns are presented \cite{mccarley2007neurobiology}. In wakefulness state, especially during closed eyes resting state (RS), there is a significant relationship between intrinsic brain activity and cognitive processes such as working memory, with SQ from a neurophysiological perspective \cite{komarov2020associations}. However, the neurophysiological characteristics of SQ in both sleep and wakefulness states, and their impact on brain function, remain unclear \cite{salehinejad2022sleep}. Thus, examining the inherent EEG features of both states can illuminate the multidimensional nature of SQ.

In this study, we employ a comprehensive neurophysiological approach to analyze SQ, considering both sleep and wakefulness states. First, we use the PSQI to divide individuals into good sleepers (GS) and poor sleepers (PS). Second, we analyze the spectrogram and wPLI to select EEG characteristics related to amplitude and phase. Next, building on the characteristics identified in these analyses, we calculate PAC to investigate complex neural patterns related to SQ and further evaluate its correlation with SQ. Finally, we apply a machine learning approach such as a support vector machine (SVM), linear discriminant analysis (LDA), $k$-nearest neighbor ($k$-NN), and logistic regression (LR) to classify SQ at an individual level, leveraging significant features identified in each EEG analysis. Our approach aims to integrate subjective sleep experiences with objective neurophysiological data, unveiling the fundamental neural characteristics determining SQ.

The main contributions of this study are as follows. \textit{i}) We investigated the roles of phase and amplitude as key EEG components in assessing SQ across sessions. \textit{ii}) We provided new neurophysiological insights by focusing on PAC, revealing how complex interactions between brain rhythms contribute to the overall quality of sleep. \textit{iii}) We demonstrated the superiority of PAC over other EEG features by applying various machine learning approaches in classifying SQ at the individual level.

\chapter{Methods}
\section{Participants}
We recruited 24 healthy individuals (13 females; mean age: 25.46 $\pm$ 2.40 years) with no known history of sleep, neurological, or psychiatric disorders. The day before the experiment, participants were advised to sleep for at least six hours and to avoid caffeine, alcohol, strenuous activity, and central nervous system-affecting medication. The Korea University Institutional Review Board (KUIRB-2021-0155-03) approved this research in alignment with the Declaration of Helsinki. Written consent was secured from all participants.

\begin{figure*}[t!]
\centering
\scriptsize
\includegraphics[width=\textwidth]{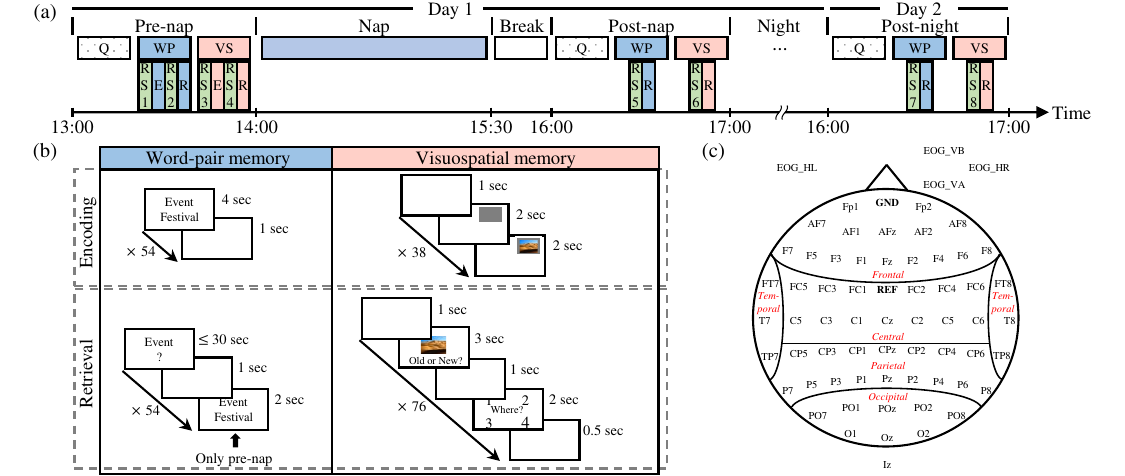}
\caption{Experimental setup. (a) The experiment consists of four main sessions: pre-nap, a 90-minute nap, post-nap, and post-night. The questionnaire (Q) component, comprising the Stanford Sleepiness Scale and the Brunel Mood Scale, is conducted alongside wakefulness, including working memory tasks and 5-minute eyes-closed resting state (RS). The RSs comprise pre-nap (RS 1 to RS 4), post-nap (RS 5 to RS 6), and post-night (RS 7 to RS 8). (b) Working memory tasks are composed of encoding (E) and retrieval (R) phases. The word-pair (WP) memory task consists of 54 trials, while the visuospatial (VS) memory task includes 38 trials for the encoding phase and 76 trials for the retrieval phase. (c) Electrode placement for the 60-channel EEG and 4-channel EOG is grouped into five brain regions: frontal, central, temporal, parietal, and occipital.}
\end{figure*}

\section{Experimental Protocol}
In a within-participants design (Figure 2.1(a)), the study included four sessions: pre-nap, nap, post-nap, and post-night. A week prior, participants completed the PSQI to assess SQ. On the experiment day, they arrived around noon, prepared for electrophysiological recordings, and completed initial questionnaires on sleepiness and mood. The experimental procedure began with a 5-minute eyes-closed RS 1 (baseline), followed by the encoding phase of the word-pair memory task. After RS 2, the retrieval phase commenced. The sequence of RS 3, encoding, RS 4, and retrieval was repeated for the visuospatial memory task. Participants then took a monitored 90-minute nap, after which they had a 30-minute break to alleviate potential sleep inertia effects. In the post-nap session, participants first completed the same questionnaires, then proceeded to the word-pair task retrieval after RS 5, followed by the visuospatial task retrieval after RS 6. A similar pattern was followed in the post-night session, with questionnaire completion preceding the retrieval phases for the word-pair and visuospatial tasks after RS 7 and RS 8, respectively. 

Participants engaged in two declarative memory tasks using Psychtoolbox as detailed in Figure 2.1(b) \cite{shin2020assessment}. For the word-pair memory task, they memorized 54 semantically related word-pairs displayed in random order during the encoding phase \cite{marshall2006boosting}. In the retrieval phase, participants were required to type their responses when cued with one word from each pair. For the visuospatial memory task including picture and location memories, participants memorized 38 images from the SUN database and their locations during the encoding phase \cite{xiao2010sun}. During the retrieval phase, participants distinguished between 38 old images they had seen before and 38 new images, also recalling the original locations of the old images. Performance was evaluated based on accuracy and reaction time for both tasks, with allowances for minor errors.

To evaluate factors related to subjective SQ, we utilized three questionnaires: the PSQI, the Stanford Sleepiness Scale (SSS), and the Brunel Mood Scale (BRUMS). The PSQI is the most common way to measure SQ and evaluate overall sleep conditions over the past month \cite{buysse1989pittsburgh}. It consists of seven components (subjective SQ, sleep latency, sleep duration, habitual sleep efficiency, sleep disturbance, use of sleep medication, and daytime dysfunction) and is calculated as a global PSQI score ranging from 0 to 21 (good sleepers (GS): $\leq$ 5; poor sleepers (PS): $>$ 5) \cite{buysse1991quantification}. The SSS is an introspective measure of the current state of sleepiness level, evaluated on a scale from 1 to 7, with higher values indicating more severe sleepiness \cite{maclean1992psychometric}. The scale ranges from 1 (feeling active, vital, alert, or wide awake) to 7 (no longer fighting sleep, sleep onset soon; having dream-like thoughts). The BRUMS is a self-reported emotional state measure consisting of 32 items \cite{terry2003construct}; after evaluating each emotion on a 5-point Likert-type scale, sets of four items were grouped to represent eight factors (anger, tension, depression, energy, fatigue, confusion, happiness, and calmness).

For subsequent analyses, participants were classified as GS (\textit{n} = 11) or PS (\textit{n} = 13) based on their PSQI scores \cite{zhou2020prediction}.

\section{Data Acquisition and Pre-processing}
EEG recordings were obtained using 60-channel Ag/AgCl electrodes according to the 10-20 international system, as illustrated in Figure 2.1(c) using BrainAmp (Brain Products GmbH, Munich, Germany). Reference and ground electrodes were situated at FCz and Fpz locations, respectively. To measure eye movements, four Ag/AgCl electrodes were placed. All electrodes were digitized with a sampling rate of 1,000 Hz and the impedance was kept below 20 k$\Omega$.

The pre-processing was done with MATLAB 2023b using the EEGLAB toolbox \cite{delorme2004eeglab} and BCILAB toolbox \cite{kothe2013bcilab}. Data were down‑sampled to 250 Hz and band‑pass filtered between 0.5 and 30 Hz using a zero‑phase finite‑impulse‑response filter with a Hamming window. Bad channels were automatically identified and removed using the \textit{pop\_rejchan} function with the kurtosis criterion \cite{wang2020novel}. To remove eye artifacts, we employed independent component analysis (ICA); components whose absolute correlation with any electrooculography channel exceeded \textit{r} = 0.70 were discarded. Additionally, we applied re-reference to the average reference and Surface Laplacian to eliminate noise.

To enhance transparency and quantify preprocessing quality, we tracked three objective indicators for each state (pre‑nap, nap, post‑nap, and post‑night): the number of bad channels removed, the count of ICA components removed, and data retention (\%). Because no epoch rejection was applied, data retention reflects only the effect of channel removal and was calculated as:

\begin{equation}
\text{Retention (\%)} = 100 \times \left( \frac{\text{Number of channels retained}}{\text{Total channels recorded}} \right)
\end{equation}

This metric quantifies the proportion of spatial EEG information retained, and numerical summaries for all three indicators are provided in Table 2.1.

\begin{table}[t!]
\caption{EEG‑preprocessing quality metrics (mean ± SEM across participants).}
\resizebox{\columnwidth}{!}{
\tiny
\renewcommand{\arraystretch}{1.0}
\begin{tabular}{@{\extracolsep{\fill}\quad}llccc}
\hline
Session    & State  & \begin{tabular}[c]{@{}c@{}}Bad   channels \\ removed (n)\end{tabular} & Data   retention (\%) & \begin{tabular}[c]{@{}c@{}}ICA   components \\ removed (n)\end{tabular} \\ \hline
Nap        & Nap    & 4.29 ± 2.13                                                           & 92.85 ± 4.66          & 0.54 ± 0.50                                                              \\
Pre-nap    & RS   1 & 3.79 ± 2.07                                                           & 93.68 ± 3.14          & 0.33 ± 0.64                                                              \\
           & RS   2 & 3.50 ± 1.57                                                           & 94.17 ± 2.62          & 0.42 ± 0.70                                                              \\
           & RS   3 & 3.88 ± 1.34                                                           & 93.53 ± 2.23          & 0.46 ± 0.68                                                              \\
           & RS   4 & 3.25 ± 1.50                                                           & 94.58 ± 2.25          & 0.54 ± 0.81                                                              \\
Post-nap   & RS   5 & 3.71 ± 1.65                                                           & 93.82 ± 2.75          & 0.46 ± 0.67                                                              \\
           & RS   6 & 3.96 ± 1.79                                                           & 93.40 ± 2.98          & 0.63 ± 0.75                                                              \\
Post-night & RS   7 & 3.13 ± 1.72                                                           & 94.80 ± 2.87          & 0.79 ± 0.86                                                              \\
           & RS   8 & 3.83 ± 1.87                                                           & 93.58 ± 2.98          & 0.75 ± 0.80          \\ \hline                                                   
\end{tabular}
}
\end{table}

The EEGs measured during sleep were segmented every 30 seconds according to the guidelines by the American Academy of Sleep Medicine and sleep experts labeled them into the five sleep stages (wake, REM, N1, N2, and N3) \cite{berry2012aasm}. The number of participants (GS/PS) in each stage was as follows: wake (10/11), REM (2/4), N1 (11/13), N2 (11/13), and N3 (7/9).

\section{Feature Estimation}
This study aimed to identify EEG differences between GS and PS across pre-nap, nap, post-nap, and post-night sessions. EEG channels were grouped into five regions of interest (ROIs) based on their scalp placement: frontal, central, temporal, parietal, and occipital (Figure 2.1(c)). These ROIs were analyzed for frequency-specific characteristics in the delta (0.5–4 Hz), theta (4-8 Hz), alpha (8–11 Hz), and beta (11–30 Hz) bands.

\subsection{Spectrogram}
We calculated the time-dependent power spectrum of the measured EEG during both sleep and wakefulness states. This analysis aimed to identify the amplitude and scalp distribution of frequencies within the EEG oscillations, revealing the regions of activation across various states \cite{park2014assessment}. For each electrode, a short-time Fourier transform was calculated from the preprocessed EEG signals using a 99\% overlapping Kaiser window over a frequency range of 0.5-30 Hz. Electrodes were grouped based on ROIs, allowing us to compute the average power for each frequency. For each participant, the analysis resulted in 20 features derived from the combination of five ROIs and four frequency bands.

\subsection{Weighted Phase Lag Index}
We estimated the functional connectivity of complex brain networks by calculating wPLI, which identifies non-zero phase lag statistical interdependencies between EEG signals from pairs of different channels. This method is robust in reducing the effects of volume conduction and provides a more reliable relationship than true phase consistency even under high signal-to-noise ratios \cite{kalafatovich2022learning}. The wPLI values, ranging from 0 to 1, indicate phase synchronization levels between channels, with 0 signifying no synchronization and 1 indicating perfect synchronization. We averaged trials for each participant to create a two-dimensional channel matrix, then averaged these within each ROI to produce 5 $\times$ 5 matrices for analysis. Considering the symmetry of the connectivity matrices, 15 unique connections among the five ROIs were selected. Combined with four frequency bands, this resulted in 60 features for each participant.

\subsection{Phase-Amplitude Coupling}
We analyzed interactions within specific ROIs, focusing on frequencies with a significant spectrogram and wPLI findings. PAC was estimated using the modulation index (MI), which is based on Kullback–Leibler (KL) divergence and Shannon entropy \cite{tort2008dynamic}, to quantify synchronization between EEG frequencies and its association with SQ. To ensure robustness, surrogate data were generated via cyclic permutation (\textit{r} = 200). For each phase bin of the phase-providing frequency, the average amplitude of the amplitude-providing frequency was calculated and normalized.

\begin{equation}
    p(j) =  \frac{A_{f_A \emptyset f_P}(j)}{\sum_{k=1}^N A_{f_A \emptyset f_P}(k)} 
\end{equation}

\noindent
where $N$ is the number of phase bins and $A_{f_A \emptyset f_P}(j)$ is the mean $f_A$ amplitude signal in the phase bin $j$ of the phase signal $\emptyset f_P$. We divided the phase into 18 bins at intervals of 20 degrees: 

\begin{equation} 
D_{KL}(P,Q) = \sum_{j=1}^{N} P(j) \log \frac{P(j)}{Q(j)} 
\end{equation}

\noindent
where $D_{KL}$ is the KL divergence distance, $P$ is the observed phase-amplitude probability density function, $Q$ is the uniform distribution, and $N$ is the number of phase bins. MI is the KL divergence divided by $\log N$ as follows:  
\begin{equation}
MI = \frac{D_{KL} (P,Q)}{\log N} 
\end{equation}

\noindent
Higher MI values signify greater non-uniformity in the distribution of phase-adjusted amplitude, indicating stronger neuronal coupling intensity. The PAC was calculated across 10 amplitude-providing frequencies and 10 phase-providing frequencies, resulting in a 10 $\times$ 10 dimension matrix.

\begin{figure*}[t!]
\centering
\scriptsize
\includegraphics[width=\textwidth]{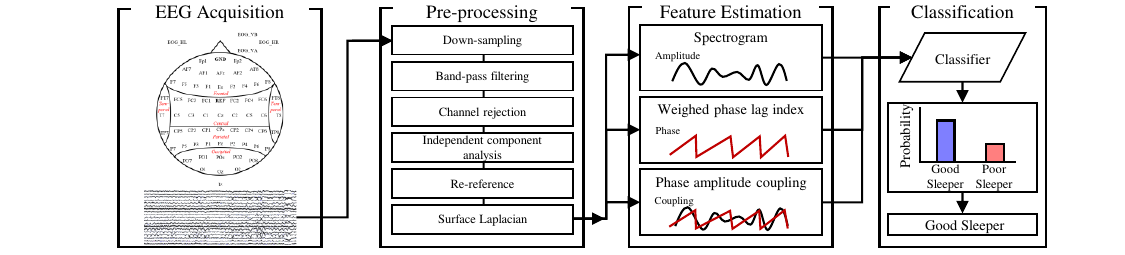}
\caption{An overall framework for classifying sleep quality based on EEG features by sessions: pre-nap, nap, post-nap, and post-night.}
\end{figure*}

\section{Classification}
We assessed SQ at an individual level by applying machine learning (Figure 2.2). To select significant spectrogram and wPLI, group-level statistical analyses were conducted across all participants to identify significant ROI and frequency band features. We averaged significant PAC dimensions that were selected by group-level statistical analyses. In the nap session, we found significant EEG features in N3. However, 16 participants were used to assess SQ because 8 participants (GS = 4) had no N3 data. Details of the input dimensions are provided in Table 2.2.

\begin{table}[t!]
\caption{Input dimensions for EEG across sleep and wakefulness sessions.}
\resizebox{\columnwidth}{!}{
\tiny
\renewcommand{\arraystretch}{1.0}
\begin{tabular}{@{\extracolsep{\fill}\quad}llc}
\hline
Session    & Feature          & Input   Dimension (N $\times$ D) \\ \hline
Nap      & Beta   power     & 16 $\times$ 2                \\
           & Delta   weighted phase lag index     & 16 $\times$ 2                \\
           & Delta-beta phase-amplitude coupling & 16 $\times$ 4                \\
Pre-nap    & Beta   power     & 24 $\times$ 3                \\
           & Delta   weighted phase lag index     & 24 $\times$ 4                \\
           & Delta-beta   phase-amplitude coupling & 24 $\times$ 12               \\
Post-nap   & Beta   power     & 24 $\times$ 4                \\
           & Delta   weighted phase lag index     & 24 $\times$ 1                \\
           & Delta-beta   phase-amplitude coupling & 24 $\times$ 4                \\
Post-night & Beta   power     & 24 $\times$ 1                \\
           & Delta   weighted phase lag index     & 24 $\times$ 2                \\
           & Delta-beta   phase-amplitude coupling & 24 $\times$ 2                \\ \hline
\end{tabular}
}
\end{table}

The identified features were standardized and inputted into classifiers, such as SVM, LDA, $k$-NN, and LR. These classifiers, having been proven effective in various pattern recognition tasks \cite{maeng2012nighttime, lee1990translation}, were chosen for their suitability in handling linearly separable data and their simpler, more interpretable models compared to more complex algorithms \cite{liu2022alterations}. Specifically, we employed SVM, training a linear kernel with L2 regularization ($\lambda = 0.1$) to balance the trade-off between margin maximization and classification error minimization \cite{amin2017classification}. We also used LDA with a diagonal linear discriminant type, which projects data to maximize class separability and reduce the risk of overfitting \cite{nkengfack2020eeg}. Additionally, we applied $k$-NN, setting $k = 5$ with inverse distance weighting to prioritize closer neighbors \cite{lee2021decoding}. Finally, we used LR, training a model with L2 regularization ($\lambda = 0.1$) to estimate class probabilities and improve generalization \cite{liu2022alterations}.

Classification was conducted for each session and feature to quantify SQ based on the session. A total of 12 classifications per classifier were performed because there were three features (beta power, delta wPLI, delta-beta PAC) and four sessions (nap, pre-nap, post-nap, post-night).

\section{Performance Metrics}
To validate the robustness of our model and prevent overfitting, we employed a leave-one-subject-out cross-validation strategy, iteratively training the model on all subjects except one and using the remaining subject as the test case in each iteration. This approach addresses our sample size limitations and provides a reliable assessment of model performance for each subject. To evaluate classifier effectiveness, we focused on accuracy (ACC), F1-score, and Cohen’s kappa.Accuracy (ACC) was calculated as:

\begin{equation} 
ACC = \frac{TP + TN}{TP + TN + FP + FN} 
\end{equation}

\noindent
where $TP$, $TN$, $FP$, and $FN$ are the number of true positives, true negatives, false positives, and false negatives.

The F1-score emphasizes the balance between false positives and false negatives (as the harmonic mean of precision and recall), making it suitable for imbalanced datasets. It was calculated as:

\begin{equation} 
F1\text{-}score = 2 \times \frac{TP}{2TP + FP + FN} 
\end{equation}

Cohen’s kappa adjusts for chance agreement, making it useful for evaluating classification performance beyond simple accuracy. It was calculated as:

\begin{equation} 
Kappa\ value = \frac{p_o-p_e}{1-p_e} = 1 - \frac{1-p_o}{1-p_e}
\end{equation}

\noindent
where $p_0$ denotes the accuracy of predictions, and $p_e$ denotes the probability of coincidence between true and predicted labels.

Together, these metrics provide a comprehensive evaluation of the model in distinguishing between GS and PS.

\section{Statistical Analysis}
To distinguish the characteristics of GS and PS, we performed statistical verification. We used the chi-square test for categorical variables and the Mann-Whitney U test for numerical variables. First, we compared demographic data and the results of SSS and BRUMS between the two groups. Second, we calculated the sleep architectures based on labeled EEG and then compared them between the two groups. Subsequently, we compared the EEG characteristics (spectrogram, wPLI, and PAC) between the two groups during pre-nap, nap, post-nap, and post-night sessions using a non-parametric permutation test (\textit{r} = 1,000). Multiple comparisons were controlled using the Benjamini-Hochberg procedure to control the false discovery rate (FDR) correction \cite{benjamini1995controlling}. Finally, associations between SQ and PAC were examined using Pearson’s correlation, followed by partial correlations controlling for age and sex, and mediation was evaluated with Sobel’s test. For all statistical analyses, the alpha level was set at 0.05. 

\chapter{Results}
\section{Physiological and Behavioral Difference by Groups} 
Based on PSQI scores, participants were categorized into GS and PS, revealing a significant difference (\textit{t} = -4.711, \textit{p} \textless{} 0.001). No significant differences were found in sex, age, or body mass index between the groups. Sleep architecture analysis showed that PS had significantly longer REM duration compared to GS (\textit{t} = -3.586, \textit{p} \textless{} 0.001). Other metrics, including the proportion of NREM sleep, total sleep time, sleep onset latency, wake after sleep onset, and sleep efficiency, showed no significant differences between the groups (Table 3.1). Questionnaire responses were also compared between the groups. The SSS, collected during the pre-nap, post-nap, and post-night sessions, showed no significant changes in subjective sleepiness levels between the two groups (Table 3.2). Similarly, BRUMS results, which assessed various emotional factors, revealed no significant group differences across all sessions (Table 3.3).

In working memory tasks, GS demonstrated significantly faster reaction times than PS in the picture memory task during the post-nap (\textit{t} = -2.195, \textit{p} = 0.026) and post-night (\textit{t} = -2.651, \textit{p} = 0.010) sessions; however, no significant differences were observed in accuracy between the two groups. On the other hand, other working memory tasks did not show significant differences in either accuracy or reaction time (Figure 3.1).

\begin{sidewaystable}[t!]
\caption{Demographics and sleep architecture between good and poor sleepers (lower quartile-upper quartile).}
\centering
\small
\renewcommand{\arraystretch}{1.0}
\begin{tabular*}{\textwidth}{@{\extracolsep{\fill}\quad}lcccc}
\hline
                                                                                  & \begin{tabular}[c]{@{}c@{}}Good sleeper \\ (\textit{n} = 11)\end{tabular} & \begin{tabular}[c]{@{}c@{}}Poor sleeper \\ (\textit{n} = 13)\end{tabular} & \begin{tabular}[c]{@{}c@{}}Total N\end{tabular} & \textit{p}-value          \\ \hline
Demographics                                                                      &                                                                 &                                                                 &                                                   &                  \\
\hspace{3mm}Sex (female, \%)                                                                  & \begin{tabular}[c]{@{}c@{}}5 (45.45)\end{tabular}            & \begin{tabular}[c]{@{}c@{}}8 (61.54)\end{tabular}             & 24                                                & 0.989            \\
\hspace{3mm}Age                                                                               & \begin{tabular}[c]{@{}c@{}}25.0 {[}25.0-26.8{]}\end{tabular} & \begin{tabular}[c]{@{}c@{}}26.0 {[}24.0-27.3{]}\end{tabular}  & 24                                                & 0.690            \\
\hspace{3mm}Body mass index (\(\text{kg/m}^2\))                                                & \begin{tabular}[c]{@{}c@{}}20.8 {[}19.3-24.5{]}\end{tabular}     & \begin{tabular}[c]{@{}c@{}}21.9 {[}18.3-25.0{]}\end{tabular}     & 24                                                & 0.562 \\
Pittsburgh sleep quality index                                                   & \begin{tabular}[c]{@{}c@{}}4.0 {[}3.0-4.0{]}\end{tabular}     & \begin{tabular}[c]{@{}c@{}}8.0 {[}6.0-8.3{]}\end{tabular}     & 24                                                & \textbf{\textless{}0.001} \\
Sleep   architecture                                                              &                                                                 &                                                                 &                                                   &                  \\
\hspace{3mm}\begin{tabular}[c]{@{}l@{}}Total time sleep (min)\end{tabular}          & \begin{tabular}[c]{@{}c@{}}75.0 {[}65.9-79.6{]}\end{tabular}  & \begin{tabular}[c]{@{}c@{}}76.0 {[}55.9-85.6{]}\end{tabular}  & 24                                                & 0.916            \\
\hspace{3mm}\begin{tabular}[c]{@{}l@{}}Sleep onset latency (min)\end{tabular}            & \begin{tabular}[c]{@{}c@{}}6.5 {[}5.5-6.5{]}\end{tabular}     & \begin{tabular}[c]{@{}c@{}}4.0 {[}2.9-8.6{]}\end{tabular}     & 24                                                & 0.686            \\
\hspace{3mm}\begin{tabular}[c]{@{}l@{}}Wake after sleep onset (min)\end{tabular}           & \begin{tabular}[c]{@{}c@{}}10.0 {[}4.0-17.1{]}\end{tabular}   & \begin{tabular}[c]{@{}c@{}}10.5 {[}2.0-25.3{]}\end{tabular}   & 24                                                & 0.936            \\
\hspace{3mm}Sleep efficiency (\%)                                                           & \begin{tabular}[c]{@{}c@{}}83.3 {[}73.2-88.5{]}\end{tabular}  & \begin{tabular}[c]{@{}c@{}}84.4 {[}62.1-95.1{]}\end{tabular}  & 24                                                & 0.896            \\
\hspace{3mm}N1 (\% of TST)                                                                & \begin{tabular}[c]{@{}c@{}}23.8 {[}13.8-37.1{]}\end{tabular}  & \begin{tabular}[c]{@{}c@{}}23.3 {[}10.7-54.1{]}\end{tabular}  & 24                                                & 0.796            \\
\hspace{3mm}N2   (\% of TST)                                                              & \begin{tabular}[c]{@{}c@{}}60.8 {[}41.1-72.2{]}\end{tabular}  & \begin{tabular}[c]{@{}c@{}}45.4 {[}37.5-66.8{]}\end{tabular}  & 24                                                & 0.578            \\
\hspace{3mm}N3 (\%   of TST)                                                              & \begin{tabular}[c]{@{}c@{}}14.4 {[}5.2-38.1{]}\end{tabular}   & \begin{tabular}[c]{@{}c@{}}10.0 {[}5.7-19.8{]}\end{tabular}  & 16                                                & 0.596            \\
\hspace{3mm}REM (\%   of TST)                                                                 & \begin{tabular}[c]{@{}c@{}}4.8 {[}2.1-7.5{]}\end{tabular}     & \begin{tabular}[c]{@{}c@{}}18.7 {[}14.1-21.9{]}\end{tabular}   & 6                                                 & \textbf{\textless{}0.001}            \\  
\hline
\end{tabular*}
\end{sidewaystable}
\clearpage

\begin{table}[t!]
\caption{Statistical difference in the SSS between good and poor sleepers pre-nap, post-nap, and post-night sessions (lower quartile-upper quartile).} 
\resizebox{\columnwidth}{!}{
\tiny
\renewcommand{\arraystretch}{1.0}
\begin{tabular}{@{\extracolsep{\fill}\quad}lcccc}
\hline
           & Good sleepers (\textit{n} = 11) & Poor sleepers (\textit{n} = 13) & Total N & \textit{p}-value \\ \hline  
Pre-nap    & 3.0 {[}2.0-3.0{]} & 3.0 {[}2.0-3.0{]} & 24      & 0.262   \\
Post-nap   & 2.0 {[}1.3-3.8{]} & 2.0 {[}2.0-3.0{]} & 24      & 0.896   \\
Post-night & 3.0 {[}2.0-3.0{]} & 3.0 {[}2.0-3.0{]} & 24      & 0.486   \\ \hline          
\end{tabular}
}
\end{table}

\begin{table}[t!]
\caption{Statistical results in the BRUMS between good and poor sleepers pre-nap, post-nap, and post-night sessions.} 
\resizebox{\columnwidth}{!}{
\tiny
\renewcommand{\arraystretch}{1.0}
\begin{tabular}{@{\extracolsep{\fill}\quad}lcccccc}
\hline
BRUMS      & \multicolumn{2}{c}{Pre-nap} & \multicolumn{2}{c}{Post-nap} & \multicolumn{2}{c}{Post-night} \\ \cline{2-7} 
           & \textit{t}-value      & \textit{p}-value      & \textit{t}-value       & \textit{p}-value      & \textit{t}-value        & \textit{p}-value       \\ \hline
Anger      & -1.010       & 0.329        & -0.877        & 0.343        & -0.471         & 0.667         \\
Tension    & -0.999       & 0.375        & -0.830        & 0.431        & -0.349         & 0.715         \\
Depression & -0.756       & 0.568        & -0.542        & 0.718        & -0.268         & 0.813         \\
Vigor      & -0.910       & 0.369        & -0.999        & 0.319        & -0.894         & 0.243         \\
Fatigue    & -0.882       & 0.395        & 0.199         & 0.844        & -0.182         & 0.231         \\
Confusion  & -0.527       & 0.666        & -0.373        & 0.726        & -0.457         & 0.463         \\
Happy      & -0.392       & 0.708        & -1.132        & 0.261        & -0.221         & 0.235         \\
Calmness   & 0.060        & 0.990        & 0.155         & 0.924        & 0.368          & 0.531         \\ \hline
\end{tabular}
}
\end{table}

\begin{figure*}[t!]
\centering
\scriptsize
\includegraphics[width=\textwidth]{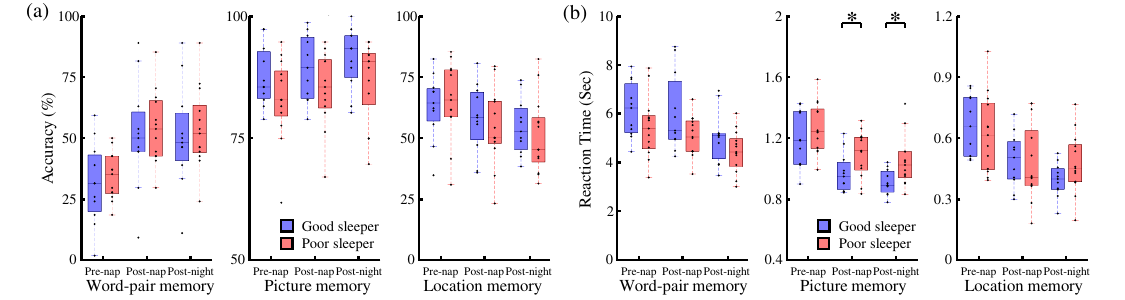}
\caption{Performance on (a) accuracy and (b) reaction time in three memory tasks for good and poor sleepers across the pre-nap, post-nap, and post-night sessions. A black asterisk indicates statistical significance (\textit{p} $<$ 0.05, FDR corrected).}
\end{figure*}

\section{Variations in Spectrogram by Sleep Quality}
We investigated spectrogram differences between GS and PS, focusing on frequency variations across ROIs during sleep and wakefulness (Figure 3.2). During the wake stage, PS had higher brain activation in all regions except the occipital region compared to GS. In the REM stage, PS exhibited overall higher activation specifically in the frontal and temporal regions. As the depth of non-REM sleep increased from N1 to N3, there was a relative increase in brain activation across overall brain regions in GS. It is noteworthy that in the N3 stage, a significant disparity was observed in the beta band, characterized by increased activity in the frontal and temporal regions for GS. No significant differences were found in other sleep stages. 

\begin{figure*}[t!]
\centering
\scriptsize
\includegraphics[width=\textwidth]{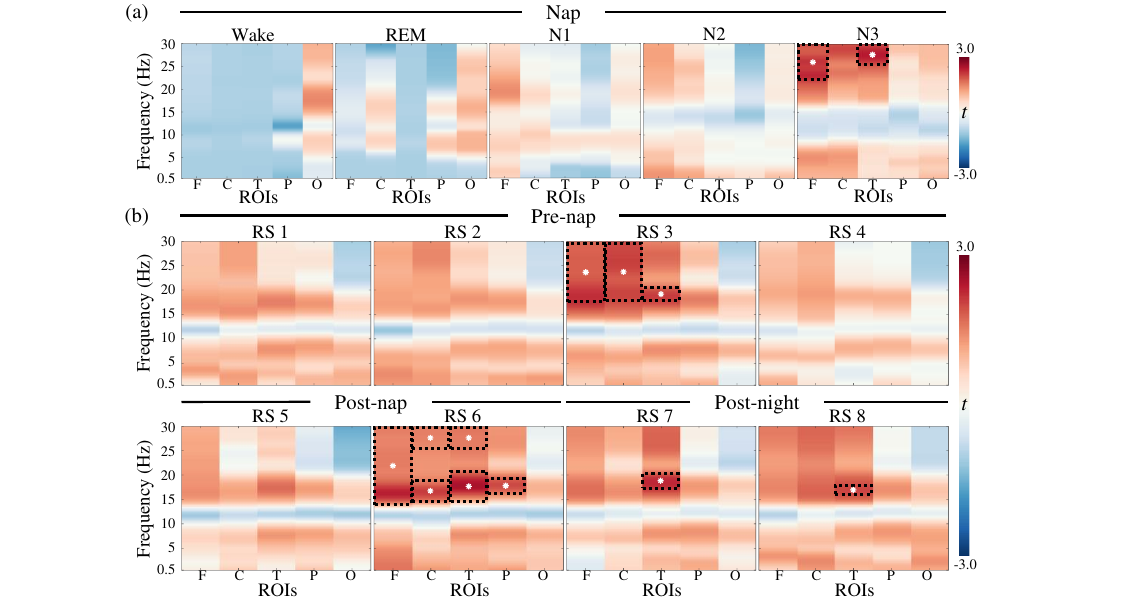}
\caption{Statistical \textit{t}-value maps showing the relative increase and decrease in spectrogram between two groups across (a) sleep stages and (b) wakefulness. This analysis highlights the frequency differences in the range of 0.5-30 Hz. Channels corresponding to the regions of interest (ROIs) – frontal, central, temporal, parietal, and occipital – are averaged. The black dotted boxes and white asterisks indicate significant areas (\textit{p} $<$ 0.05, FDR corrected).}
\end{figure*}

In the RS analysis, GS generally exhibited greater brain activity than PS in all brain regions during RSs. Distinct statistical differences between the two groups were evident in RS 3, RS 6, RS 7, and RS 8, particularly in the beta band. In RS 3, marked differences were observed in the frontal, central, and temporal regions, while RS 6 exhibited notable differences across all brain regions except the occipital region. RS 7 and RS 8 showed the most prominent differences in the temporal region. Conversely, no significant differences were found at RS 1, indicating comparable initial spectral activity between GS and PS. Other RSs and frequency bands revealed no additional effects.

In addition, quantitative ROI–based analyses of the beta band–region combinations across both sleep and RSs confirmed that GS exhibited greater spectral power than PS in the regions showing significant group differences (Figure 3.3).

\begin{figure*}[t!]
\centering
\scriptsize
\includegraphics[width=\textwidth]{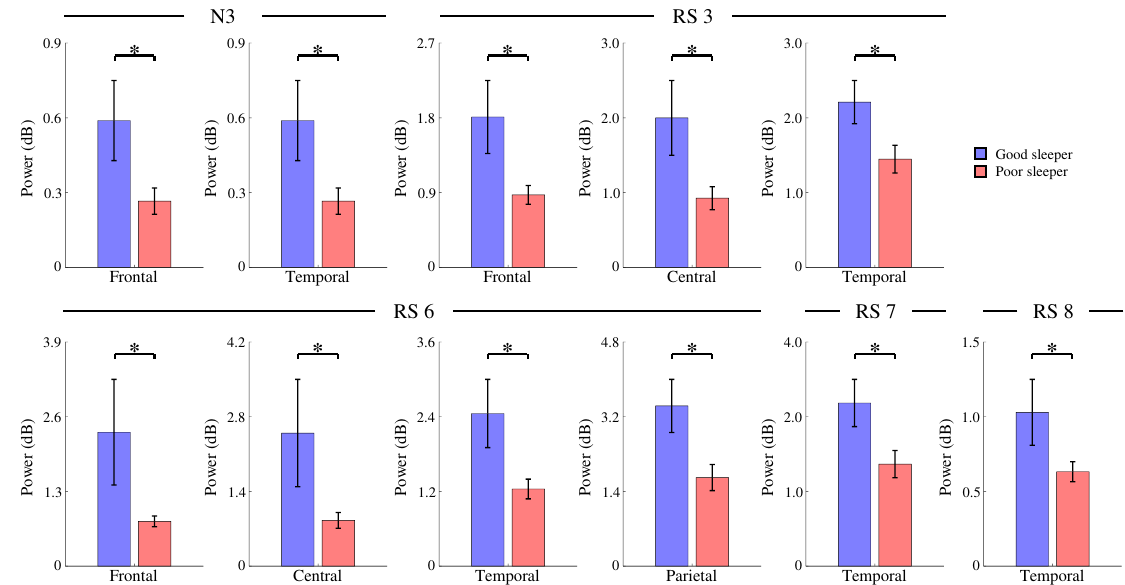}
\caption{Statistical \textit{t}-value maps showing the relative increase and decrease in spectrogram between two groups across (a) sleep stages and (b) wakefulness. This analysis highlights the frequency differences in the range of 0.5-30 Hz. Channels corresponding to the regions of interest (ROIs) – frontal, central, temporal, parietal, and occipital – are averaged. The black dotted boxes and white asterisks indicate significant areas (\textit{p} $<$ 0.05, FDR corrected).}
\end{figure*}

\section{Cerebral Synchronization of Sleep and Wakefulness}
We computed the wPLI during sleep and wakefulness to compare brain connectivity between GS and PS. Figure 3.4(a) shows A significant difference in wPLI across various sleep stages. In the N2 stage, significant differences were observed in the delta wPLI across all brain regions except the frontal, frontocentral, frontotemporal, central, and centrotemporal regions. Additionally, a notable difference in the alpha wPLI was found in the frontal region. In the N3 stage, GS exhibited significantly higher connectivity in the delta wPLI than PS, notably in the temporal and parietal regions. No significant differences were noted in other sleep stages and frequency bands. 

\begin{figure*}[t!]
\centering
\scriptsize
\includegraphics[width=\textwidth]{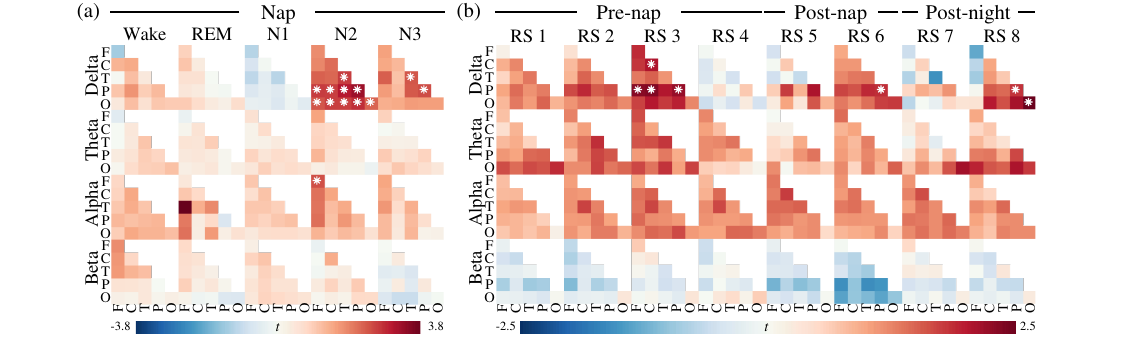}
\caption{Results of statistical analyses of weighted phase lag index between two groups across (a) sleep stages and (b) wakefulness, focusing on brain regions: frontal, central, temporal, parietal, and occipital. Comparisons are made across various frequency bands, including delta, theta, alpha, and beta bands. The white asterisk represents significant connectivity between brain regions (\textit{p} $<$ 0.05, FDR corrected).}
\end{figure*}

During RSs in pre-nap, post-nap, and post-night sessions (Figure 3.4(b), we compared the connectivity between ROIs for each frequency band between the two groups. GS exhibited higher connectivity in delta, theta, and alpha wPLI, while PS had higher connectivity in beta wPLI. In the delta wPLI, GS showed statistically higher connectivity across certain RSs. Specifically, in RS 3, GS had higher connectivity in the frontoparietal, central, centroparietal, and parietal regions. During RS 6, GS demonstrated higher connectivity in the parietal region. For RS 8, GS had higher connectivity in the parietal and occipital regions. On the other hand, RS 1 did not yield any notable group differences, implying that baseline functional connectivity patterns were similar between groups. No further group-level effects emerged in the remaining RSs or frequency bands.

\section{Phase–Amplitude Coupling Patterns Reflecting Sleep Quality} 
We utilized significant EEG characteristics to calculate the exchange of information between the phase of low-frequency and the amplitude of high-frequency during sleep and wakefulness. In the N3 stage, our findings highlighted a marked interaction between the phase of delta oscillations and the amplitude of beta oscillations (Figure 3.5(a)). GS exhibited a significantly higher MI, evident in pronounced delta-beta PAC, with particularly strong coupling between delta-phase in temporal and parietal regions and beta-amplitude in frontal and temporal regions—suggesting enhanced functional coordination across these cortical networks in GS. 

\begin{figure*}[t!]
\centering
\scriptsize
\includegraphics[width=\textwidth]{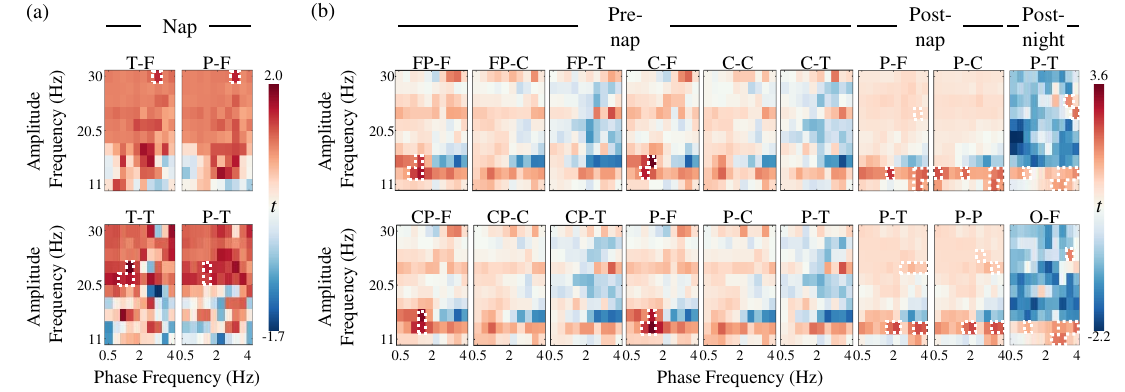}
\caption{Comodulograms illustrating delta-beta phase-amplitude coupling across different conditions: (a) sleep, specifically the N3 stage, and (b) wakefulness, with a specific focus on pre-nap (RS 3), post-nap (RS 6), and post-night (RS 8). These comodulograms illustrate the statistical difference in averaged modulation index values between two groups, highlighting significant coupling between phase in the delta band and amplitude in the beta band in significant brain regions. Regions of significant coupling are indicated by white dotted boxes (\textit{p} $<$ 0.05, FDR corrected).}
\end{figure*}

We focused on the statistical differences in MI values between groups during significant RS sessions: pre-nap (RS 3), post-nap (RS 6), and post-night (RS 8), as illustrated in Figure 3.5(b). For RS 3, the interaction between the phase of delta oscillations and the amplitude of beta oscillations was highlighted. In this context, GS exhibited significantly higher MI values specifically for the coupling of frontal beta amplitude with delta phase in the frontoparietal, central, centroparietal, and parietal regions. By contrast, MI between beta amplitude in the central or temporal regions and delta phase showed no significant group differences. For RS 6, the analysis revealed that GS exhibited significantly higher MI values in relation to beta amplitudes in the frontal, central, temporal, and parietal regions, particularly with the delta phase in the parietal region. Furthermore, for RS 8, GS demonstrated statistically significant coupling related to beta amplitudes in the temporal region, especially with the delta phase in the parietal and occipital regions. These findings underline the robust delta-beta PAC in GS, suggesting a complex neural ensemble underpinning SQ.

We further investigated the significant correlation between delta-beta PAC and PSQI scores. This relationship was consistent during nap, as shown in Figure 3.6(a), and in wakefulness, including pre-nap, post-nap, and post-night, as shown in Figure 3.6(b). Higher SQ was consistently associated with stronger delta-beta PAC, which was negatively correlated with PSQI scores. After adjusting for age and sex, partial correlations between delta-beta PAC and PSQI scores remained significant across both sleep and wakefulness states, indicating that the associations are robust to demographic influences. Mediation analyses using Sobel tests revealed no significant indirect effects of age (\textit{z} = 0.945, \textit{p} = 0.345) or sex (\textit{z} = 0.703, \textit{p} = 0.482) on PSQI via PAC, suggesting that delta-beta PAC does not mediate demographic influences on SQ. These findings suggest that robust delta-beta PAC may serve as a neurophysiological marker of SQ, providing insights into the distinct neural dynamics associated with SQ.

\begin{figure*}[t!]
\centering
\scriptsize
\includegraphics[width=\textwidth]{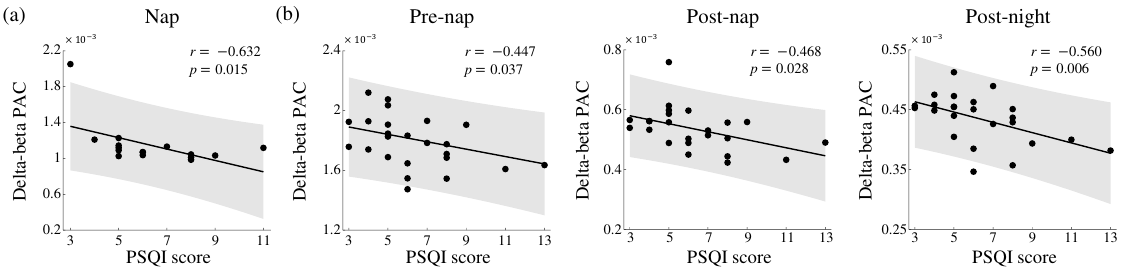}
\caption{Correlation between delta-beta PAC and PSQI score across different conditions: (a) N3 stage and (b) wakefulness conditions including pre-nap (RS 3), post-nap (RS 6), and post-night (RS 8). Gray shading represents the 95\% confidence intervals around the regression line.}
\end{figure*}

\section{Classification Performance of EEG Features in Sleep Quality} 
We evaluated the performance of key EEG features—beta power, delta wPLI, and delta-beta PAC—during the nap, pre-nap, post-nap, and post-night sessions to classify SQ at the individual level. These features were inputted into the SVM, LDA, $k$-NN, and LR classifiers, selected for their potential to capture different aspects of SQ. As shown in Table 3.4, delta-beta PAC generally achieved the highest classification performance across most classifiers and sessions, indicating its robustness as a marker of SQ. For example, in the nap session, delta-beta PAC with the SVM classifier achieved an ACC of 0.81, an F1-score of 0.73, and a Cohen’s kappa of 0.60. Similarly, in the post-night session, delta-beta PAC with the LR classifier attained an ACC of 0.75, an F1-score of 0.75, and a Cohen’s kappa of 0.50. In contrast, beta power and delta wPLI generally yielded lower performance across most classifiers and sessions. For instance, beta power with SVM during the nap session showed an ACC of 0.69, an F1-score of 0.55, and a Cohen’s kappa of 0.33, reflecting limited discriminative ability. While some instances, such as delta wPLI with $k$-NN in the pre-nap session (ACC of 0.88, F1-score of 0.87, and Cohen’s kappa of 0.75), showed high values, these were less consistent across sessions. Overall, these results underscore the utility of delta-beta PAC in reliably distinguishing SQ across various states, with notable performance improvements over other EEG features in most cases. This evaluation highlights the capability of delta-beta PAC to capture nuanced patterns of SQ and further emphasizes the complexities involved in using neural markers for accurate classification. 

\begin{sidewaystable}[t!]
\caption{Classification results of significant EEG features of two groups according to sleep and wakefulness}
\centering
\small
\renewcommand{\arraystretch}{1.0}
\begin{tabular*}{\textwidth}{@{\extracolsep{\fill}\quad}llcccccccccccc}
\hline
\multirow{2}{*}{Classifier} & \multirow{2}{*}{Feature} & \multicolumn{3}{c}{Nap} & \multicolumn{3}{c}{Pre-nap} & \multicolumn{3}{c}{Post-nap} & \multicolumn{3}{c}{Post-night} \\ \cline{3-14}
                            &                          & ACC    & F1    & Kappa  & ACC     & F1      & Kappa   & ACC     & F1      & Kappa    & ACC      & F1       & Kappa    \\ \hline
\multirow{3}{*}{SVM}        & Beta   power             & 0.69   & 0.55  & 0.33   & 0.67    & 0.60    & 0.32    & 0.71    & 0.63    & 0.40     & \textbf{0.75}     & 0.67     & 0.48     \\
                            & Delta   wPLI             & 0.63   & 0.57  & 0.24  & 0.63    & 0.40    & 0.21    & 0.63    & 0.40    & 0.21     & 0.67     & 0.56     & 0.31    \\
                            & Delta-beta   PAC         & \textbf{0.81}   & \textbf{0.73}  & \textbf{0.60}   & \textbf{0.75}    & \textbf{0.73}    & \textbf{0.50}    & \textbf{0.75}    & \textbf{0.73}    & \textbf{0.50}     & \textbf{0.75}     & \textbf{0.77}     & \textbf{0.51}     \\ \hline
\multirow{3}{*}{LDA}        & Beta   power             & 0.63   & 0.40  & 0.19   & 0.71    & 0.63    & 0.40    & 0.71    & 0.63    & 0.40     & 0.67     & 0.56     & 0.31     \\
                            & Delta   wPLI             & 0.63   & 0.57  & 0.24   & 0.63    & 0.40    & 0.21    & 0.63    & 0.40    & 0.21     & 0.58     & 0.44     & 0.14     \\
                            & Delta-beta   PAC         & \textbf{0.75}   & \textbf{0.60}  & \textbf{0.46}   & \textbf{0.75}    & \textbf{0.72}    & \textbf{0.50}   & \textbf{0.75}    & \textbf{0.72}    & \textbf{0.50}     & \textbf{0.75}     & \textbf{0.75}     & \textbf{0.50}     \\ \hline
\multirow{3}{*}{$k$-NN}       & Beta   power             & 0.63   & 0.50  & 0.21   & 0.67    & 0.50    & 0.30    & 0.71    & 0.63    & 0.40     & 0.67     & 0.56     & 0.31    \\
                            & Delta   wPLI             & 0.50   & 0.33  & -0.05  & \textbf{0.83}    & \textbf{0.82}    & \textbf{0.66}    & 0.50    & 0.33    & -0.04    & 0.67     & 0.69     & 0.35     \\
                            & Delta-beta   PAC         & \textbf{0.75}   & \textbf{0.60}  & \textbf{0.46}   & 0.58    & 0.55    & 0.16    & \textbf{0.79}    & \textbf{0.78}    & \textbf{0.58}     & \textbf{0.71}     & \textbf{0.72}    & \textbf{0.42}     \\ \hline
\multirow{3}{*}{LR}         & Beta   power             & 0.56   & 0.36  & 0.07   & 0.67    & 0.56    & 0.31    & 0.67    & 0.56    & 0.31     & 0.67     & 0.56     & 0.31     \\
                            & Delta   wPLI             & 0.56   & 0.46  & 0.10   & 0.58    & 0.38    & 0.12    & 0.58    & 0.38    & 0.12    & 0.58     & 0.44     & 0.14     \\
                            & Delta-beta   PAC         & \textbf{0.75}   & \textbf{0.60}  & \textbf{0.46}   & \textbf{0.71}    & \textbf{0.67}    & \textbf{0.41}    & \textbf{0.71}    & \textbf{0.67}    & \textbf{0.41}     & \textbf{0.75}     & \textbf{0.75}     & \textbf{0.50}  \\ \hline
\end{tabular*}
\end{sidewaystable}
\clearpage

Furthermore, we evaluated classifiers for other significant EEG features across different states, as shown in Table 3.5. The analysis revealed that delta wPLI and alpha wPLI during N2 sleep, and beta power during RS 7 yielded moderate overall performance. These results underscore the potential of our models to classify SQ based on specific EEG markers in distinct sleep and RS conditions, but also further emphasize the complexities and inherent challenges involved in interpreting neural signals to achieve accurate assessments of SQ.

\begin{table}[t!]
\caption{Classification results of significant EEG features of two groups according to state}
\resizebox{\columnwidth}{!}{
\tiny
\renewcommand{\arraystretch}{1.2}
\begin{tabular}{@{\extracolsep{\fill}\quad}lllccc}
\hline
Classifier           & State & Feature      & ACC  & F1   & Kappa \\ \hline
\multirow{3}{*}{SVM} & N2    & Delta   weighted phase lag index & 0.79 & 0.76 & 0.58  \\
                     & N2    & Alpha   weighted phase lag index & 0.67 & 0.63 & 0.33  \\
                     & RS 7   & Beta   power & 0.67 & 0.60 & 0.32  \\ \hline
\multirow{3}{*}{LDA} & N2    & Delta   weighted phase lag index & 0.79 & 0.76 & 0.58  \\
                     & N2    & Alpha   weighted phase lag index & 0.58 & 0.50 & 0.15  \\
                     & RS 7   & Beta   power & 0.63 & 0.57 & 0.24  \\ \hline
\multirow{3}{*}{$k$-NN} & N2    & Delta   weighted phase lag index & 0.79 & 0.76 & 0.58  \\
                     & N2    & Alpha   weighted phase lag index & 0.58 & 0.50 & 0.15  \\
                     & RS 7   & Beta   power & 0.58 & 0.50 & 0.15  \\ \hline
\multirow{3}{*}{LR}  & N2    & Delta   weighted phase lag index & 0.67 & 0.60 & 0.32  \\
                     & N2    & Alpha   weighted phase lag index & 0.58 & 0.50 & 0.15  \\
                     & RS 7   & Beta   power & 0.67 & 0.60 & 0.32  \\ \hline
\end{tabular}
}
\end{table}

\chapter{Discussion}
This study aimed to assess SQ in sleep and wakefulness using subjective and objective measures. We observed that GS exhibited a significantly faster reaction time in the picture memory task compared to PS during both post-nap and post-night sessions, linking SQ to cognitive performance. In EEG analysis, we identified statistically significant differences in beta power, delta wPLI, and delta-beta PAC between the two groups. Distinct patterns in these EEG features, particularly evident during N3 and specific RSs, underscore the relationship between various aspects of brain activity and the overall quality of sleep. Leveraging these significant EEG features, we employed machine learning approaches to classify SQ at an individual level. Notably, the performance of delta-beta PAC emerged as the most robust indicator, highlighting its potential as a key marker in SQ assessment.

\section{Influence of Sleep Quality on Demographics, Sleep Architecture, and Cognitive Performance}
Our study investigated the intricate interactions between physiological and behavioral factors influencing SQ. As shown in Table 3.1, no significant differences were found in most aspects of sleep architecture related to SQ between the two groups, which may be attributed to individual variations in sleep patterns \cite{yetton2018quantifying}. However, a notable exception was the duration of REM sleep, where PS exhibited significantly longer periods. This suggests a possible compensatory mechanism or an altered need for REM sleep in PS, hinting at deeper physiological or psychological differences \cite{beersma1990rem}. Conversely, analysis of questionnaire responses revealed no significant differences in subjective sleepiness and emotions between the groups, potentially due to methodological constraints, such as limited questionnaire scoring ranges, which might not capture subtle physiological variations \cite{zavecz2020relationship}. 

Furthermore, the impact of SQ on cognitive performance was found to be task-specific. GS showed superior reaction times in picture memory task during both post-nap and post-night sessions, indicating more efficient information processing and storage. This finding aligns with previous research demonstrating that high-quality sleep enhances memory consolidation and cognitive processing, particularly for tasks involving picture memory \cite{mednick2003sleep}. According to another study, sleep facilitates neural processes essential for encoding and retrieving complex information in tasks requiring focused attention and memory \cite{gais2004declarative}. These processes contribute to faster response times and improved task performance in well-rested individuals. In contrast, word-pair and location memory tasks did not show significant performance differences, suggesting these cognitive functions may be less influenced by SQ \cite{zavecz2020relationship}. These findings highlight the multifaceted role of sleep and its broad impact on health and behavior, covering sleep architecture, subjective experiences, and cognitive functions.

\section{Distinct Beta Activity in Sleep and Wakefulness}
Our study revealed distinct differences in beta power between GS and PS, as evidenced by spectrogram analysis during N3 and RSs. Specifically, in GS, increased beta activity was observed in the frontal and temporal regions during N3, which diverges from the expected decrease in high-frequency oscillations typically seen in deep sleep \cite{kang2022difference}. Previous research has reported that it may be involved in sleep-dependent learning processes in which the brain reorganizes and integrates information \cite{mednick2003sleep}. This pattern, potentially linked to a nap after working memory tasks, may imply a unique aspect of brain function in GS, possibly reflecting enhanced memory consolidation and cognitive processing during N3 stage. 

In addition, our RS analysis revealed that GS exhibit a distinctive pattern of increased beta activity in RS 3, RS 6, RS 7, and RS 8, with a notable emphasis on RS 3, RS 6, and RS 8 after the retrieval phase of working memory tasks. This enhanced beta activity may indicate a higher level of cognitive processing, suggesting that GS exhibit superior capabilities in assimilating and integrating complex cognitive information compared to PS \cite{heister2013resting}. This observation suggests that beta power observed across RSs likely reflects the distinct cognitive demands and contexts of each RS. For instance, RS 3, conducted after memory encoding and retrieval phases, may support early consolidation, RS 6, following a nap, likely reflects reorganization and integration, while RS 7 and RS 8, following repeated cognitive tasks and nighttime sleep, display patterns of readiness and recovery, indicating adaptation to varying cognitive demands. Such patterns suggest that brain activities during RSs reflect cognitive processing capabilities, potentially contributing to efficient memory integration \cite{li2020impact}.

\section{Delta Connectivity Across Sleep and Wakefulness by Sleep Quality}
The current study suggests that GS have significantly higher brain synchronization during sleep and wakefulness. In the N2 stage, GS showed higher delta wPLI and alpha wPLI across various brain regions, indicative of potentially more active or efficient cognitive processing networks. These results are consistent with existing literature findings, highlighting weak brain connectivity and synchronization loss in PS have been noted \cite{leistedt2009altered}. Similarly, in the N3 stage, the presence of higher delta wPLI among different brain regions in GS underscores the critical role of delta band in deep, restorative sleep, suggesting more effective neural integration that potentially facilitates improved SQ and its associated restorative benefits. This is consistent with research linking poor SQ to decreased brain synchronization in the delta band during deep sleep \cite{desjardins2017eeg}. 

In RSs, distinct wPLI patterns emerged, reflecting specific neural connectivity adaptations unique to each RS. For example, RS 3 shows increased connectivity across memory-related regions, likely supporting initial memory consolidation following cognitive tasks. RS 6, conducted after a nap, exhibits broader connectivity that may reflect neural reorganization and recovery, while RS 8, conducted after nighttime sleep, shows connectivity patterns that suggest neural readiness and stability, preparing the brain for subsequent cognitive demands. These variations in delta wPLI across RSs align with the idea that GS maintain a more synchronized and efficient network, indicative of enhanced neural integration, stability, and cognitive readiness. This finding complements evidence that poor SQ adversely affects brain structural and functional networks, influencing cognitive functions and overall brain health \cite{lian2023reduced}. Collectively, these insights establish clear distinctions in brain connectivity related to SQ, shedding light on its influence on neural interactions.

\section{Impact on Sleep Quality through Neural Dynamics of Delta-Beta PAC}
Delta-beta PAC emerges as a key finding in our study, significantly influencing both sleep and RS. Particularly in the N3 stage, we observed an increased MI value in the phase of the temporal, frontotemporal, and parietal regions, and in the amplitude of the frontal and temporal regions among GS. This suggests an intensified neurophysiological engagement in these regions is crucial for deep sleep and associated with enhanced cognitive processing and memory consolidation \cite{weber2021coupling}. Consistent with this, an overnight sleep study observed robust delta-beta PAC during NREM sleep, and its close alignment with our N3-stage results demonstrates that our 90-minute nap paradigm reliably captures this conserved mechanism of sleep-quality–dependent neural coordination \cite{cox2019heterogeneous}. 

Elevated MI values of delta-beta PAC in various brain regions among GS during RS 3, RS 6, and RS 8 point to active neurophysiological processes following cognitive engagement. These distinct PAC patterns across RSs likely reflect the specific cognitive and rest contexts associated with each session. For instance, RS 3, conducted after memory retrieval, shows strong delta-beta PAC in memory-related regions, supporting early consolidation processes. RS 6, following a nap, displays PAC patterns that may indicate neural reorganization and integration following rest, while RS 8, conducted after nighttime sleep, demonstrates PAC patterns suggesting neural readiness and adaptation for upcoming cognitive demands. This finding suggests that neural patterns in sleep have parallels in the RS after cognitive exertion \cite{cox2019heterogeneous}. Moreover, sleep-restriction research shows that this coupling weakens with insufficient sleep \cite{zhang2023phase}, mirroring our finding that individuals with PS exhibit reduced MI values. This convergence of findings across pre-nap, nap, post-nap, and post-night sessions further reinforces the external validity of our nap-based paradigm. Extending these observations to clinical populations, clinical studies have shown significantly reduced delta-beta PAC in individuals with insomnia disorder compared to healthy controls \cite{guo2023effects} and in patients with depressive disorders \cite{venanzi2024delta}, both characterized by pronounced sleep disturbances. These convergent findings support the utility of delta-beta PAC as a translational biomarker of sleep disruption across both sleep disorders and related psychiatric conditions.

Furthermore, our analysis revealed a significant negative correlation between delta-beta PAC and PSQI scores, suggesting that stronger delta-beta coupling may reflect enhanced neurophysiological support for SQ. This finding aligns with the observed role of delta-beta PAC in facilitating cognitive processing and memory consolidation. These insights highlight the influence of SQ on PAC dynamics, shedding light on the neural mechanisms underlying SQ.

\section{Interpreting EEG Features for Sleep Quality Classification by Sessions}
In classifying individual SQ based on EEG features, our findings demonstrated the strong performance of delta-beta PAC compared to beta power and delta wPLI. Delta-beta PAC consistently achieved the highest score across multiple metrics, underscoring its potential as a key indicator for assessing SQ. This superior performance likely arises from the unique ability of PAC to capture critical interactions between amplitude and phase in EEG signals \cite{shi2018cross}. Additionally, the SVM classifier proved particularly effective among the classifiers tested, showing reliable performance in distinguishing SQ across both sleep and wakefulness states, driven by the predictive strength of delta-beta PAC \cite{samiee2017time}. Our analysis across pre-nap, nap, post-nap, and post-night sessions highlights the unique benefits of classifying SQ in each state. The pre-nap session, serving as an RS baseline, may offer valuable information about readiness for sleep, capturing initial neural patterns associated with relaxation and the transition into sleep \cite{larson2011modulation}. The nap session, specifically during N3 sleep, provides insight into deep sleep characteristics tied to restorative sleep processes and memory consolidation \cite{weber2021coupling}. The post-nap session reflects the immediate impact of the nap on neural recovery and transition back into wakefulness, offering a perspective on the effectiveness of sleep in enhancing cognitive readiness \cite{milner2009benefits}. Finally, the post-night session captures the longer-term effects of overnight sleep, highlighting residual neural patterns that may correlate with subjective SQ and next-day functioning \cite{diekelmann2010memory}. These insights suggest that each session captures distinct aspects of SQ, making a multi-session approach valuable for comprehensive assessment. Overall, our findings support a machine learning-based approach to EEG analysis, providing deeper insights into the neurological mechanisms of sleep health. The consistent performance of delta-beta PAC across sessions underscores its value as a reliable marker of SQ.

\section{Limitations \& Future Works}
Our study has several limitations. We included only healthy individuals, which limits clinical generalizability, so future work will recruit participants with sleep disorders such as insomnia and obstructive sleep apnea as well as psychiatric conditions including major depressive disorder and schizophrenia to validate delta-beta PAC as a neurophysiological marker of impaired SQ in both research and clinical settings. To enable direct quantitative comparison with typical overnight sleep, future work will involve collecting overnight EEG recordings under identical experimental protocols and SQ assessments. Our 60-channel EEG system limits spatial resolution and may not fully mitigate volume-conduction effects. Therefore, in future studies, we will employ high-density EEG recordings to enable finer mapping of oscillatory and connectivity patterns across cortical regions. To further enhance signal quality, we plan to apply advanced modeling techniques, such as uncertainty-aware decoders used for high-fidelity reconstruction \cite{lee2020uncertainty}, which can improve the robustness of source localization. Moreover, we aim to develop a multimodal sleep monitoring system that integrates EEG with non-contact video analysis. Computer vision methodologies for human body pose reconstruction \cite{yang2007reconstruction} and human action recognition \cite{ahmad2006human} could be adapted to quantify sleep postures and movements. Integrating these with automatic gesture recognition frameworks would provide a comprehensive understanding of sleep behavior beyond neurophysiology \cite{hwang2006full}. Furthermore, to provide a more nuanced understanding of the relationship between sleep and cognitive performance, we will include EEG analyses related to memory \cite{shin2021predicting}. This approach will help elucidate the broader cognitive implications of SQ, thereby significantly enhancing the scope of our research.

\chapter{Conclusion}
Our study investigated the neurological mechanisms of SQ across various states of sleep and wakefulness, employing sophisticated methods such as spectrogram, wPLI, and PAC analysis using EEG. We identified pronounced differences in the interaction between delta-phase and beta-amplitude, which were notably distinct between individuals with GS and PS. Additionally, utilizing a machine learning approach in classifying SQ individually, PAC analysis significantly outperformed other methods in terms of accuracy, demonstrating statistically significant superiority. These findings not only deepen our understanding of the neurological basis of SQ but also establish delta-beta PAC as a reliable marker for distinguishing between GS and PS across sleep and wakefulness states. By identifying unique PAC patterns associated with SQ in pre-nap, nap, post-nap, and post-night sessions, delta-beta PAC provides an effective tool for assessing individual sleep health in clinical and diverse contexts.\\

\newpage
\renewcommand\bibname{Reference}
\bibliographystyle{ieeetr}
\bibliography{REFERENCE}

@article{delorme2004eeglab,
  title={{EEGLAB}: {A}n open source toolbox for analysis of single-trial {EEG} dynamics including independent component analysis},
  author={A. Delorme and S. Makeig},
  journal={J. Neurosci. Methods},
  volume={134},
  number={1},
  pages={9--21},
  year={2004},
}

@article{kothe2013bcilab,
  title={{BCILAB}: {A} platform for brain-computer interface development},
  author={C. A. Kothe and S. Makeig},
  journal={J. Neural Eng.},
  volume={10},
  number={5},
  pages={056014},
  year={2013},
  publisher={IOP Publishing}
}

@article{pilcher1997sleep,
  title={Sleep quality versus sleep quantity: {R}elationships between sleep and measures of health, well-being and sleepiness in college students},
  author={Pilcher, June J and Ginter, Douglas R and Sadowsky, Brigitte},
  journal={J. Psychosomat. Res.},
  volume={42},
  number={6},
  pages={583--596},
  year={1997},
  publisher={Elsevier}
}

@article{lee2020frontal,
  title={Frontal {EEG} asymmetry of emotion for the same auditory stimulus},
  author={M. Lee and {\mbox{G.-H}}. Shin and {\mbox{S.-W}}. Lee},
  journal={IEEE Access},
  volume={8},
  pages={107200--107213},
  year={2020},
  publisher={IEEE}
}

@article{tort2008dynamic,
  title={Dynamic cross-frequency couplings of local field potential oscillations in rat striatum and hippocampus during performance of a {T}-maze task},
  author={Adriano B L Tort and Mark A Kramer and Catherine Thorn and Daniel J Gibson and Yasuo Kubota and Ann M Graybiel and Nancy J Kopell},
  journal={Proc. Natl. Acad. Sci. U. S. A.}, 
  volume={105},
  number={51},
  pages={20517--20522},
  year={2008},
  publisher={National Acad Sciences}
}

@article{samiee2017time,
  title={Time-resolved phase-amplitude coupling in neural oscillations},
  author={Samiee, Soheila and Baillet, Sylvain},
  journal={Neuroimage},
  volume={159},
  pages={270--279},
  year={2017},
  publisher={Elsevier}
}

@article{bakhshayesh2019detecting,
  title={Detecting synchrony in {EEG}: {A} comparative study of functional connectivity measures},
  author={Bakhshayesh, Hanieh and Fitzgibbon, Sean P and Janani, Azin S and Grummett, Tyler S and Pope, Kenneth J},
  journal={Comput. Biol. Med.},
  volume={105},
  pages={1--15},
  year={2019},
  publisher={Elsevier}
}

@article{buysse1989pittsburgh,
  title={The {P}ittsburgh {S}leep {Q}uality {I}ndex: {A} new instrument for psychiatric practice and research},
  author={Buysse, Daniel J and Reynolds III, Charles F and Monk, Timothy H and Berman, Susan R and Kupfer, David J},
  journal={Psychiatry Res.},
  volume={28},
  number={2},
  pages={193--213},
  year={1989},
  publisher={Elsevier}
}

@article{maclean1992psychometric,
  title={Psychometric evaluation of the {S}tanford {S}leepiness {S}cale},
  author={Maclean, Alistair W and Fekken, G Cynthia and Saskin, Paul and Knowles, John B},
  journal={J. Sleep Res.},
  volume={1},
  number={1},
  pages={35--39},
  year={1992},
  publisher={Wiley Online Library}
}

@article{leistedt2009altered,
  title={Altered sleep brain functional connectivity in acutely depressed patients},
  author={Leistedt, Samuel JJ and Coumans, Nathalie and Dumont, Martine and Lanquart, Jean-Pol and Stam, Cornelis J and Linkowski, Paul},
  journal={Hum. Brain Mapp.},
  volume={30},
  number={7},
  pages={2207--2219},
  year={2009},
  publisher={Wiley Online Library}
}

@article{berry2012aasm,
  title={The {AASM} manual for the scoring of sleep and associated events},
  author={Berry, Richard B and Brooks, Rita and Gamaldo, Charlene E and Harding, Susan M and Marcus, C and Vaughn, Bradley V},
  journal={Rules, Terminology and Technical Specifications, Darien, Illinois, American Academy of Sleep Medicine},
  volume={176},
  year={2012}
}

@article{ramlee2017sways,
  title={What sways people’s judgment of sleep quality? {A} quantitative choice-making study with good and poor sleepers},
  author={Ramlee, Fatanah and Sanborn, Adam N and Tang, Nicole KY},
  journal={Sleep},
  volume={40},
  number={7},
  pages={zsx091},
  year={2017},
  publisher={Oxford University Press}
}

@article{zavecz2020relationship,
  title={The relationship between subjective sleep quality and cognitive performance in healthy young adults: {E}vidence from three empirical studies},
  author={Zavecz, Zs{\'o}fia and Nagy, Tam{\'a}s and Galk{\'o}, Adrienn and Nemeth, Dezso and Janacsek, Karolina},
  journal={Sci. Rep.},
  volume={10},
  number={1},
  pages={1--12},
  year={2020},
  publisher={Nature Publishing Group}
}

@article{lendner2020electrophysiological,
  title={An electrophysiological marker of arousal level in humans},
  author={Lendner, Janna D and Helfrich, Randolph F and Mander, Bryce A and Romundstad, Luis and Lin, Jack J and Walker, Matthew P and Larsson, Pal G and Knight, Robert T},
  journal={eLife},
  volume={9},
  pages={e55092},
  year={2020},
  publisher={eLife Sciences Publications, Ltd}
}

@article{shi2018cross,
  title={Cross-frequency transfer entropy characterize coupling of interacting nonlinear oscillators in complex systems},
  author={Shi, Wenbin and Yeh, Chien-Hung and Hong, Yang},
  journal={IEEE Trans. Biomed. Eng.},
  volume={66},
  number={2},
  pages={521--529},
  year={2018},
  publisher={IEEE}
}

@article{zhang2023phase,
  title={Phase-amplitude coupling of {G}o/{N}ogo task-related neuronal oscillation decreases for humans with insufficient sleep},
  author={Zhang, Peng and Sun, Chuancai and Liu, Zhongqi and Zhou, Qianxiang},
  journal={Sleep},
  volume={46},
  number={11},
  pages={zsad243},
  year={2023},
  publisher={Oxford University Press US}
}

@article{komarov2020associations,
  title={Associations among emotional state, sleep quality, and resting-state {EEG} spectra: {A} longitudinal study in graduate students},
  author={Komarov, Oleksii and Ko, Li-Wei and Jung, Tzyy-Ping},
  journal={IEEE Trans. Neural Syst. Rehabil. Eng.},
  volume={28},
  number={4},
  pages={795--804},
  year={2020},
  publisher={IEEE}
}

@article{mccarter2022physiological,
  title={Physiological Markers of Sleep Quality: {A} Scoping Review},
  author={McCarter, Stuart J and Hagen, Philip T and Louis, Erik K St and Rieck, Thomas M and Haider, Clifton R and Holmes, David R and Morgenthaler, Timothy I},
  journal={Sleep Med. Rev.},
  pages={101657},
  year={2022},
  publisher={Elsevier}
}

@article{desjardins2017eeg,
  title={{EEG} functional connectivity prior to sleepwalking: {E}vidence of interplay between sleep and wakefulness},
  author={Desjardins, Marie-{\`E}ve and Carrier, Julie and Lina, Jean-Marc and Fortin, Maxime and Gosselin, Nadia and Montplaisir, Jacques and Zadra, Antonio},
  journal={Sleep},
  volume={40},
  number={4},
  pages={zsx024},
  year={2017},
  publisher={Oxford University Press US}
}

@article{kalafatovich2022learning,
  title={Learning Spatiotemporal Graph Representations for Visual Perception Using {EEG} Signals},
  author={Kalafatovich, Jenifer and Lee, Minji and Lee, Seong-Whan},
  journal={IEEE Trans. Neural Syst. Rehabil. Eng.},
  volume={31},
  pages={97--108},
  year={2022},
  publisher={IEEE}
}

@inproceedings{xiao2010sun,
  title={Sun database: {L}arge-scale scene recognition from abbey to zoo},
  author={J. Xiao and J. Hays and K. A. Ehinger and A. Oliva and A. Torralba},
  booktitle={Proc. IEEE Comput. Soc. Conf. Comput. Vis. Pattern Recognit. (CVPR)},
  pages={3485--3492},
  year={2010},
}

@inproceedings{shin2020assessment,
  title={Assessment of Unconsciousness for Memory Consolidation Using {EEG} Signals},
  author={Shin, Gi-Hwan and Lee, Minji and Lee, Seong-Whan},
  booktitle={Proc. IEEE Int. Conf. Syst., Man, and Cybern. (SMC)},
  pages={513--519},
  year={2020},
}

@article{marshall2006boosting,
  title={Boosting slow oscillations during sleep potentiates memory},
  author={L. Marshall and H. Helgad{\'o}ttir and N. M{\"o}lle and J. Born},
  journal={Nature},
  volume={444},
  number={7119},
  pages={610--613},
  year={2006},
  publisher={Nature Publishing Group}
}

@article{zhou2020prediction,
  title={Prediction and classification of sleep quality based on phase synchronization related whole-brain dynamic connectivity using resting state {fMRI}},
  author={Zhou, Zhongxing and Cai, Biao and Zhang, Gemeng and Zhang, Aiying and Calhoun, Vince D and Wang, Yu-Ping},
  journal={Neuroimage},
  volume={221},
  pages={117190},
  year={2020},
  publisher={Elsevier}
}

@article{salehinejad2022sleep,
  title={Sleep-dependent upscaled excitability, saturated neuroplasticity, and modulated cognition in the human brain},
  author={Salehinejad, Mohammad Ali and Ghanavati, Elham and Reinders, J{\"o}rg and Hengstler, Jan G and Kuo, Min-Fang and Nitsche, Michael A},
  journal={eLife},
  volume={11},
  pages={e69308},
  year={2022},
  publisher={eLife Sciences Publications Limited}
}

@article{beersma1990rem,
  title={{REM} sleep deprivation during 5 hours leads to an immediate {REM} sleep rebound and to suppression of non-{REM} sleep intensity},
  author={Beersma, DGM and Dijk, DJ and Blok, CGH and Everhardus, I},
  journal={Electroencephalogr. Clin. Neurophysiol.},
  volume={76},
  number={2},
  pages={114--122},
  year={1990},
  publisher={Elsevier}
}

@article{wang2019consistency,
  title={Consistency and dynamical changes of directional information flow in different brain states: {A} comparison of working memory and resting-state using {EEG}},
  author={Wang, Ruimin and Ge, Sheng and Zommara, Noha Mohsen and Ravienna, Karine and Espinoza, Teodora and Iramina, Keiji},
  journal={Neuroimage},
  volume={203},
  pages={116188},
  year={2019},
  publisher={Elsevier}
}

@article{mccarley2007neurobiology,
  title={Neurobiology of {REM} and {NREM} sleep},
  author={McCarley, Robert W},
  journal={Sleep Med.},
  volume={8},
  number={4},
  pages={302--330},
  year={2007},
  publisher={Elsevier}
}

@article{varela2001brainweb,
  title={The brainweb: {P}hase synchronization and large-scale integration},
  author={Varela, Francisco and Lachaux, Jean-Philippe and Rodriguez, Eugenio and Martinerie, Jacques},
  journal={Nat. Rev. Neurosci.},
  volume={2},
  number={4},
  pages={229--239},
  year={2001},
  publisher={Nature Publishing Group UK London}
}

@article{kang2022difference,
  title={Difference in spectral power density of sleep electroencephalography between individuals without insomnia and frequent hypnotic users with insomnia complaints},
  author={Kang, Jae Myeong and Cho, Seo-Eun and Moon, Jong Youn and Kim, Soo In and Kim, Jong Won and Kang, Seung-Gul},
  journal={Sci. Rep.},
  volume={12},
  number={1},
  pages={2117},
  year={2022},
  publisher={Nature Publishing Group UK London}
}

@article{weber2021coupling,
  title={Coupling of gamma band activity to sleep spindle oscillations--a combined {EEG/MEG} study},
  author={Weber, Frederik D and Supp, Gernot G and Klinzing, Jens G and M{\"o}lle, Matthias and Engel, Andreas K and Born, Jan},
  journal={Neuroimage},
  volume={224},
  pages={117452},
  year={2021},
  publisher={Elsevier}
}

@article{cox2019heterogeneous,
  title={Heterogeneous profiles of coupled sleep oscillations in human hippocampus},
  author={Cox, Roy and R{\"u}ber, Theodor and Staresina, Bernhard P and Fell, Juergen},
  journal={Neuroimage},
  volume={202},
  pages={116178},
  year={2019},
  publisher={Elsevier}
}

@article{pallos2007quality,
  title={The quality of sleep and factors associated with poor sleep in Japanese graduate students},
  author={Pallos, Henrik and Gergely, Viktor and Yamada, Naoto and Miyazaki, Soichiro and Okawa, Masako},
  journal={Sleep Biol. Rhythms},
  volume={5},
  pages={234--238},
  year={2007},
  publisher={Springer}
}

@article{heister2013resting,
  title={Resting-state neuronal oscillatory correlates of working memory performance},
  author={Heister, David and Diwakar, Mithun and Nichols, Sharon and Robb, Ashley and Angeles, Anne Marie and Tal, Omer and Harrington, Deborah L and Song, Tao and Lee, Roland R and Huang, Mingxiong},
  journal={PLoS One},
  volume={8},
  number={6},
  pages={e66820},
  year={2013},
  publisher={Public Library of Science San Francisco, USA}
}

@article{li2020impact,
  title={The impact of mental fatigue on brain activity: {A} comparative study both in resting state and task state using {EEG}},
  author={Li, Gang and Huang, Shan and Xu, Wanxiu and Jiao, Weidong and Jiang, Yonghua and Gao, Zhao and Zhang, Jianhua},
  journal={BMC Neurosci.},
  volume={21},
  pages={1--9},
  year={2020},
  publisher={Springer}
}

@article{wang2020novel,
  title={A novel method to understand neural oscillations during full-body reaching: a combined {EEG} and 3{D} virtual reality study},
  author={Wang, Wei-En and Ho, Rachel LM and Gatto, Bryan and Van Der Veen, Susanne M and Underation, Matthew K and Thomas, James S and Antony, Ajay B and Coombes, Stephen A},
  journal={IEEE Trans. Neural Syst. Rehabil. Eng.},
  volume={28},
  number={12},
  pages={3074--3082},
  year={2020},
  publisher={IEEE}
}

@article{mednick2003sleep,
  title={Sleep-dependent learning: {A} nap is as good as a night},
  author={Mednick, Sara and Nakayama, Ken and Stickgold, Robert},
  journal={Nat. Neurosci.},
  volume={6},
  number={7},
  pages={697--698},
  year={2003},
  publisher={Nature Publishing Group US New York}
}

@article{lian2023reduced,
  title={Reduced Resting-State {EEG} Power Spectra and Functional Connectivity after 24 and 36 Hours of Sleep Deprivation},
  author={Lian, Jie and Xu, Lin and Song, Tao and Peng, Ziyi and Zhang, Zheyuan and An, Xin and Chen, Shufang and Zhong, Xiao and Shao, Yongcong},
  journal={Brain Sci.},
  volume={13},
  number={6},
  pages={949},
  year={2023},
  publisher={MDPI}
}

@article{yetton2018quantifying,
  title={Quantifying sleep architecture dynamics and individual differences using big data and Bayesian networks},
  author={Yetton, Benjamin D and McDevitt, Elizabeth A and Cellini, Nicola and Shelton, Christian and Mednick, Sara C},
  journal={PLoS One},
  volume={13},
  number={4},
  pages={e0194604},
  year={2018},
  publisher={Public Library of Science San Francisco, CA USA}
}

@article{terry2003construct,
  title={Construct validity of the Profile of Mood States—{A}dolescents for use with adults},
  author={Terry, Peter C and Lane, Andrew M and Fogarty, Gerard J},
  journal={Psychol. Sport Exerc.},
  volume={4},
  number={2},
  pages={125--139},
  year={2003},
  publisher={Elsevier}
}

@article{park2014assessment,
  title={Assessment of cognitive engagement in stroke patients from single-trial {EEG} during motor rehabilitation},
  author={Park, Wanjoo and Kwon, Gyu Hyun and Kim, Da-Hye and Kim, Yun-Hee and Kim, Sung-Phil and Kim, Laehyun},
  journal={IEEE Trans. Neural Syst. Rehabil. Eng.},
  volume={23},
  number={3},
  pages={351--362},
  year={2014},
  publisher={IEEE}
}

@article{buysse1991quantification,
  title={Quantification of subjective sleep quality in healthy elderly men and women using the {P}ittsburgh {S}leep {Q}uality {I}ndex {(PSQI)}},
  author={Buysse, Daniel J and Reynolds III, Charles F and Monk, Timothy H and Hoch, Carolyn C and Yeager, Amy L and Kupfer, David J},
  journal={Sleep},
  volume={14},
  number={4},
  pages={331--338},
  year={1991},
  doi = {https://doi.org/10.1093/sleep/14.4.331},
  publisher={Oxford University Press}
}

@article{lee2021decoding,
  title={Decoding finger tapping with the affected hand in chronic stroke patients during motor imagery and execution},
  author={Lee, Minji and Jeong, Ji-Hoon and Kim, Yun-Hee and Lee, Seong-Whan},
  journal={IEEE Trans. Neural Syst. Rehabil. Eng.},
  volume={29},
  pages={1099--1109},
  year={2021},
  publisher={IEEE}
}

@article{liu2022alterations,
  title={Alterations in patients with first-episode depression in the eyes-open and eyes-closed conditions: {A} resting-state {EEG} study},
  author={Liu, Shuang and Liu, Xiaoya and Yan, Danfeng and Chen, Sitong and Liu, Yanli and Hao, Xinyu and Ou, Wenwen and Huang, Zhenni and Su, Fangyue and He, Feng and others},
  journal={IEEE Trans. Neural Syst. Rehabil. Eng.},
  volume={30},
  pages={1019--1029},
  year={2022},
  publisher={IEEE}
}

@article{amin2017classification,
  title={Classification of {EEG} signals based on pattern recognition approach},
  author={Amin, Hafeez Ullah and Mumtaz, Wajid and Subhani, Ahmad Rauf and Saad, Mohamad Naufal Mohamad and Malik, Aamir Saeed},
  journal={Front. Comput. Neurosci.},
  volume={11},
  pages={103},
  year={2017},
  publisher={Frontiers Media SA}
}

@article{nkengfack2020eeg,
  title={{EEG} signals analysis for epileptic seizures detection using polynomial transforms, linear discriminant analysis and support vector machines},
  author={Nkengfack, Laurent Chanel Djoufack and Tchiotsop, Daniel and Atangana, Romain and Louis-Door, Val{\'e}rie and Wolf, Didier},
  journal={Biomed. Signal Process. Control},
  volume={62},
  pages={102141},
  year={2020},
  publisher={Elsevier}
}

@article{garbarino2021role,
  title={Role of sleep deprivation in immune-related disease risk and outcomes},
  author={Garbarino, Sergio and Lanteri, Paola and Bragazzi, Nicola Luigi and Magnavita, Nicola and Scoditti, Egeria},
  journal={Commun. Biol.},
  volume={4},
  number={1},
  pages={1304},
  year={2021},
  publisher={Nature Publishing Group UK London}
}

@article{gais2004declarative,
  title={Declarative memory consolidation: mechanisms acting during human sleep},
  author={Gais, Steffen and Born, Jan},
  journal={Learn. Mem.},
  volume={11},
  number={6},
  pages={679--685},
  year={2004},
  publisher={Cold Spring Harbor Lab}
}

@article{larson2011modulation,
  title={Modulation of the brain’s functional network architecture in the transition from wake to sleep},
  author={Larson-Prior, Linda J and Power, Jonathan D and Vincent, Justin L and Nolan, Tracy S and Coalson, Rebecca S and Zempel, John and Snyder, Abraham Z and Schlaggar, Bradley L and Raichle, Marcus E and Petersen, Steven E},
  journal={Prog. Brain Res.},
  volume={193},
  pages={277--294},
  year={2011},
  publisher={Elsevier}
}

@article{milner2009benefits,
  title={Benefits of napping in healthy adults: impact of nap length, time of day, age, and experience with napping},
  author={Milner, Catherine E and Cote, Kimberly A},
  journal={J. Sleep Res.},
  volume={18},
  number={2},
  pages={272--281},
  year={2009},
  publisher={Wiley Online Library}
}

@article{diekelmann2010memory,
  title={The memory function of sleep},
  author={Diekelmann, Susanne and Born, Jan},
  journal={Nat. Rev. Neurosci.},
  volume={11},
  number={2},
  pages={114--126},
  year={2010},
  publisher={Nature Publishing Group UK London}
}

@article{lee2023seriessleepnet,
  title={Series{S}leep{N}et: an {EEG} time series model with partial data augmentation for automatic sleep stage scoring},
  author={Lee, Minji and Kwak, Heon-Gyu and Kim, Hyeong-Jin and Won, Dong-Ok and Lee, Seong-Whan},
  journal={Front. Psychol.},
  volume={14},
  pages={1188678},
  year={2023},
  publisher={Frontiers Media SA}
}

@article{guo2023effects,
  title={Effects on resting-state {EEG} phase-amplitude coupling in insomnia disorder patients following 1 Hz left dorsolateral prefrontal cortex r{TMS}},
  author={Guo, Yongjian and Zhao, Xumeng and Zhang, Xiaozi and Li, Minpeng and Liu, Xiaoyang and Lu, Ling and Liu, Jiayi and Li, Yan and Zhang, Shan and Yue, Lirong and others},
  journal={Hum. Brain Mapp.},
  volume={44},
  number={8},
  pages={3084--3093},
  year={2023},
  publisher={Wiley Online Library}
}

@article{venanzi2024delta,
  title={Delta-Beta Coupling in Adolescents With Depression},
  author={Venanzi, Lisa and Dickey, Lindsay and Pegg, Samantha and Kujawa, Autumn},
  journal={J. Psychophysiol.},
  year={2024},
  publisher={Hogrefe Publishing}
}

@article{benjamini1995controlling,
  title={Controlling the false discovery rate: a practical and powerful approach to multiple testing},
  author={Benjamini, Yoav and Hochberg, Yosef},
  journal={J. Roy. Stat. Soc. Ser. B (Methodol.)},
  volume={57},
  number={1},
  pages={289--300},
  year={1995},
  publisher={Wiley Online Library}
}

@inproceedings{shin2021predicting,
  title={Predicting the transition from short-term to long-term memory based on deep neural network},
  author={Shin, Gi-Hwan and Kweon, Young-Seok and Lee, Minji},
  booktitle={Int. Winter Conf. Brain-Computer Interface (BCI)},
  pages={1--5},
  year={2021},
  organization={IEEE}
}

@article{lee1990translation,
  title={Translation-, rotation-and scale-invariant recognition of hand-drawn symbols in schematic diagrams},
  author={Lee, Seong-Whan and Kim, Jin H and Groen, Frans CA},
  journal={Int. J. Pattern Recognit. Artif. Intell.},
  volume={4},
  number={01},
  pages={1--25},
  year={1990},
  publisher={World Scientific}
}

@inproceedings{maeng2012nighttime,
  title={Nighttime face recognition at long distance: {C}ross-distance and cross-spectral matching},
  author={Maeng, Hyunju and Liao, Shengcai and Kang, Dongoh and Lee, Seong-Whan and Jain, Anil K},
  booktitle={Proc. Asian Conf. Comput. Vis. (ACCV)},
  pages={708--721},
  year={2012},
  organization={Springer}
}

@article{min2022attentional,
  title={Attentional feature pyramid network for small object detection},
  author={Min, Kyungseo and Lee, Gun-Hee and Lee, Seong-Whan},
  journal={Neural Netw.},
  volume={155},
  pages={439--450},
  year={2022},
  publisher={Elsevier}
}

@inproceedings{lim2000text,
  title={Text extraction in {MPEG} compressed video for content-based indexing},
  author={Lim, Young-Kyu and Choi, Song-Ha and Lee, Seong-Whan},
  booktitle={Proc. Int. Conf. Pattern Recognit. (ICPR)},
  volume={4},
  pages={409--412},
  year={2000},
  organization={IEEE}
}

@inproceedings{lee2020uncertainty,
  title={Uncertainty-aware mesh decoder for high fidelity 3d face reconstruction},
  author={Lee, Gun-Hee and Lee, Seong-Whan},
  booktitle={Proc. IEEE Comput. Soc. Conf. Comput. Vis. Pattern Recognit. (CVPR)},
  pages={6100--6109},
  year={2020}
}

@inproceedings{ahmad2006human,
  title={Human action recognition using multi-view image sequences},
  author={Ahmad, Mohiuddin and Lee, Seong-Whan},
  booktitle={Proc. Int. Conf. Autom. Face Gesture Recognit.},
  pages={523--528},
  year={2006},
  organization={IEEE}
}

@inproceedings{hwang2006full,
  title={A full-body gesture database for automatic gesture recognition},
  author={Hwang, Bon-Woo and Kim, Sungmin and Lee, Seong-Whan},
  booktitle={Proc. Int. Conf. Autom. Face Gesture Recognit.},
  pages={243--248},
  year={2006},
  organization={IEEE}
}

@article{roh2007accurate,
  title={Accurate object contour tracking based on boundary edge selection},
  author={Roh, Myung-Cheol and Kim, Tae-Yong and Park, Jihun and Lee, Seong-Whan},
  journal={Pattern Recognit.},
  volume={40},
  number={3},
  pages={931--943},
  year={2007},
  publisher={Elsevier}
}

@article{yang2007reconstruction,
  title={Reconstruction of 3{D} human body pose from stereo image sequences based on top-down learning},
  author={Yang, Hee-Deok and Lee, Seong-Whan},
  journal={Pattern Recognit.},
  volume={40},
  number={11},
  pages={3120--3131},
  year={2007},
  publisher={Elsevier}
}

@article{lee2001automatic,
  title={Automatic video parsing using shot boundary detection and camera operation analysis},
  author={Lee, Mee-Sook and Yang, Yun-Mo and Lee, Seong-Whan},
  journal={Pattern Recognit.},
  volume={34},
  number={3},
  pages={711--719},
  year={2001},
  publisher={Elsevier}
}

\end{document}